\documentclass[amsmath,amssymb,pre,twocolumn,floatfix,superscriptaddress]{revtex4-2}
\usepackage[utf8]{inputenc}
\usepackage{color}
\usepackage{graphicx}
\usepackage{dcolumn}
\usepackage{times}
\usepackage{soul}
\usepackage{siunitx}
\usepackage{afterpage}
\usepackage{xcolor}
\usepackage{tabularx}
\usepackage{placeins}
\usepackage{bm}
\usepackage{setspace}
\usepackage{siunitx}
\usepackage{soul}

\graphicspath{{figure*s/}}

\begin{document}
	\title{Range and strength of mechanical interactions of force dipoles in elastic fiber networks}
	
	\author{Abhinav Kumar}
	\affiliation{Department of Physics, University of California, Merced, Merced, CA 95343, USA}
	\author{David A. Quint}
	\affiliation{Lawrence Livermore National Laboratory, Livermore, California 94550, USA}
	\author{Kinjal Dasbiswas}\email{kdasbiswas@ucmerced.edu}
	\affiliation{Department of Physics, University of California, Merced, Merced, CA 95343, USA}

\begin{abstract}

Mechanical forces generated by myosin II molecular motors drive diverse cellular processes, most notably shape change, division and locomotion.  These forces may be transmitted over long range through the cytoskeletal medium - a disordered, viscoelastic network of biopolymers. The resulting cell size scale force chains can in principle mediate mechanical interactions between distant actomyosin units, leading to self-organized structural order in the cell cytoskeleton. To investigate this process, we model the actin cytoskeleton on a percolated fiber lattice network, where fibers are modeled as linear elastic elements that can both bend and stretch, whereas myosin motors exert contractile force dipoles.  We quantify the range and heterogeneity of force transmission in these networks in response to a force dipole, showing how these depend on varying bond dilution and fiber bending-to-stretching stiffness ratio. By analyzing clusters of nodes connected to highly strained bonds, as well as the decay rate of strain energy with distance from the force dipole, we show that long-range force transmission is screened out by fiber bending in diluted networks.. We further characterize the difference in the propagation of tensile and compressive forces. This leads to a dependence  of the mechanical interaction between a pair of force dipoles on their mutual separation and orientation  In more homogeneous networks, the interaction between force dipoles recapitulates the power law dependence on separation distance predicted by continuum elasticity theory, while in diluted  networks, the interactions are short-ranged and fluctuate strongly with local network configurations.  Altogether, our work suggests that elastic interactions between force dipoles in disordered, fibrous media can act as an organizing principle in biological materials.

\end{abstract}

\maketitle

\section{Introduction}

Mechanical forces generated in the cytoskeleton of animal cells underlie essential functions such as cell motility \cite{gardel_10}, shape change \cite{gardel_15,lecuit_11} and cell division \cite{green_12}. These forces are exerted by molecular motors of the myosin II family which bind to and slide actin filaments \cite{hartman2012myosin}. Actin filaments, together with other semiflexible biopolymers, constitutes the cytoskeleton of the cell:  a disordered, cross-linked,  viscoelastic network of fibers that supports the generation and transmission of mechanical forces \cite{phillips}.  Cells use these forces to deform and sense their mechanical micro-environment, which potentially allows long-range mechanical communication between cells \cite{Reinhart2008, Nitsan2016}, in addition to chemical signaling. Similarly, there is evidence that myosin II motors  may sense and interact with each other through mechanical strains that they generate in the cytoskeletal medium, leading to self-organized, structures with long-range order in the cell cytoskeleton \cite{hu_17, dasbiswas_18}. A minimal, coarse-grained model for cellular force distribution at both scales, whether by individual actomyosin  units in the cytoskeleton, or by whole cells adhered to an extracellular matrix, is a contractile force dipole embedded in an elastic medium \cite{schwarz_02, safran_13}. The spatial distribution of the deformation generated by such a force dipole, for example how far strain propagates from the position of the dipole, depends on the mechanical properties of the medium. For fibrous networks such as the extracellular matrix comprising collagen or fibrin, it has been shown that the force transmission can be longer-ranged than in a linear elastic medium \cite{rudnicki2013nonlinear}. It has been suggested that the softening of fibers under compression through buckling, together with stiffening under tension, can drive the enhanced range of force transmission \cite{xu_15, Ronceray16, Sopher2018, sarkarM2022forcechains}.  How forces propagate from an active, contractile force dipole through a heterogeneous elastic medium, and how that facilitates mechanical interaction between a pair of distant force dipoles, are therefore significant biophysical questions. This is analogous to elastic interaction of defects that act as sources of stress in passive materials \cite{eshelby1956continuum}, and are found to play a role in the organization of other disordered media such as granular packings \cite{Pappu2022PhysRevE}.

At time scales too short for cytoskeletal remodeling to occur, the cytoskeleton behaves as an elastic material that can sustain and transmit mechanical stresses \cite{fabry_11}. The disordered cytoskeleton can therefore be modeled as a network of elastic fibers that resist both stretching and bending with elastic moduli, $\mu$ and $\kappa$, respectively \cite{head_03,  PicuSM2011, broedersz_14}. In the limit where the fibers are significantly shorter than the persistence length, thermal fluctuations may be ignored and an athermal, linear elastic model for the fibers may be used to describe the cytoskeletal network instead of the nonlinear, entropic elastic constitutive relations of semiflexible polymers \cite{broedersz_14}. A convenient modeling strategy to generate such an elastic network with a disordered architecture is to start from a triangular lattice and then to remove bonds at random, with a probability $p$ of bonds being present, which corresponds to a specific average coordination number, $\langle z\rangle = 6 p$ \cite{vignaud2021stress, broedersz2011criticality, das2012redundancy, broedersz_11}. The macroscopic elastic properties of such a  network depend on the single fiber mechanics as well as the network geometry, particularly the coordination number. The macroscopic response of the network to shear is controlled by the rigidity percolation threshold for central force networks, which is calculated by Maxwell to be $p^T_{CF} = 2/3$, just from constraint counting for a 2D triangular lattice \cite{maxwell1864calculation}. This has been numerically verified to be $p^N_{CF} \approx 0.66$ \cite{Jacobs1995}, below which a 2D triangular network of central force springs (\emph{i.e.} with no resistance to bending, $\kappa=0$) lose their rigidity and become completely floppy.  Such a rigidity percolation with changing coordination number is a generic phase transition in disordered elastic materials, occuring in network glasses \cite{feng2016nonlinear} and colloidal gels \cite{Zhang2019}, in addition to fiber networks.



Below the central-force isostatic limit ($p<p_{CF}$), an elastic network may be stabilized by the bending stiffness of fibers, here represented by the energy cost of changing angles between collinear bonds in the triangular lattice.   Since slender fibers are typically much softer to bend than to stretch ( $\kappa/(\mu l^{2}) \ll 1$), they would rather bend than stretch when stressed. The $p< p_{CF}$ regime allows such stretch-free deformation modes. While not completeley floppy, these soft modes are characterized by very low mechanical energy that scales with $\kappa$.  When diluted even further, a 2D triangular lattice of fibers with finite bending modulus ($\kappa>0$), exhibits the rigidity percolation transition, with shear modulus becoming $G=0$ at the bending isostatic threshold, $p_{b} \approx 0.44$ \cite{broedersz2011criticality, das2012redundancy, huisman2011internal, Zaccone2013elastic}. The rigidity percolation threshold may be further lowered by imposing additional constraints, such as bond torsion \cite{das2012redundancy}. For bending and stretching only, there is an intermediate dilution regime $p_{b} < p < p_{CF}$, where the  network response to shear is dominated by fiber bending modes, leading to a scaling of the shear modulus with the bending stiffness, $ G \sim \kappa$ \cite{wyart2005rigidity, das2012redundancy, broedersz2011criticality,  feng2016nonlinear}. The network deformations in this regime involve bond rotations along the shearing direction without stretching, and are therefore qualitatively different, from the stretching-dominated, over-coordinated ($p > p_{CF}$), where the shear modulus is much higher, $G \sim \mu$.  The nonlinear elastic properties of under-coordinated fiber networks are demonstrated by their ``stress-strain'' curves, which stiffen dramatically by orders of magnitude, as the network transitions under shear from the bending to stretching-dominated regime \cite{sharma2016strain}.  Close to this transition, the bending and stretching modes are coupled, and this bend-stretch coupled regime can be wider if the ratio of bending to stretching moduli  increases \cite{broedersz2011criticality}.   Realistic biopolymer networks, such as occurring in the cytoskeleton of  living cells or in purified actin gels, are expected to have a typical coordination number in the range of $3< z< 4$ ($0.5 <  p < 0.67 $), where $3$ and $4$ correspond to a branch point or fibers crossing, respectively  \cite{Gardel04, broedersz_14}. This puts biopolymer networks in the under-coordinated regime but close to criticality, and allows for strong stiffening response to shear. A similar stiffening transition has been shown to occur for bulk deformations \cite{ArzashPhysRevE2022}.  


While the global elastic properties of such networks including the rigidity percolation transition have been studied extensively in past works \cite{das2012redundancy, broedersz_11, broedersz_14, AlvaradoSM2017}, we explore here the response of the network to local dipole forces embedded within the network. These may represent for instance the contractile forces actively generated by myosin molecular motors bound to the actin fibers in the cytoskeletal network. In particular, we aim to investigate the mechanical interactions that may arise between a pair of such force dipoles. The mechanical response in disordered fiber networks with bending is characterized by non-affine, heterogeneous deformations, which cannot be captured by a continuum linear elastic model.  We explore the range of force transmission from a source force dipole modeling the active contraction generated by myosin motors, through a 2D disordered elastic network in various regimes. The range of force transmission has been shown to be enhanced by fiber buckling \cite{Ronceray16}, and similarly by nonlinear constitutive relations for individual fibers that may arise from buckling \cite{mann2019force}.  In contrast to these recent works \cite{Ronceray16, mann2019force, ruiz2022force} which study force transmission from large, isotropic force distributions (modeling cells in extracellular matrix), we focus on small, anisotropic force dipoles that deform networks of linear elastic fibers with stretching and bending only. We aim to study the consequences of fiber bending separately from fiber buckling. This assumption may be applicable to networks with smaller forces and thicker fibers, that make fiber buckling less likely \cite{landau_lifshitz_elasticity}.  Further, we show how a pair of active force dipoles may interact at long range through their mutual deformations of the intervening elastic network. Such interactions leading to the mutual attraction and alignment of force dipoles and may underlie the self-organization of initially disordered actomyosin units into ordered structures such as stress fibers in the cytoskeleton \cite{vignaud2021stress, Lehtimaki2021}.


\section{Model}

        \begin{figure*}[htp]
            \centering
            \includegraphics[width=16cm]{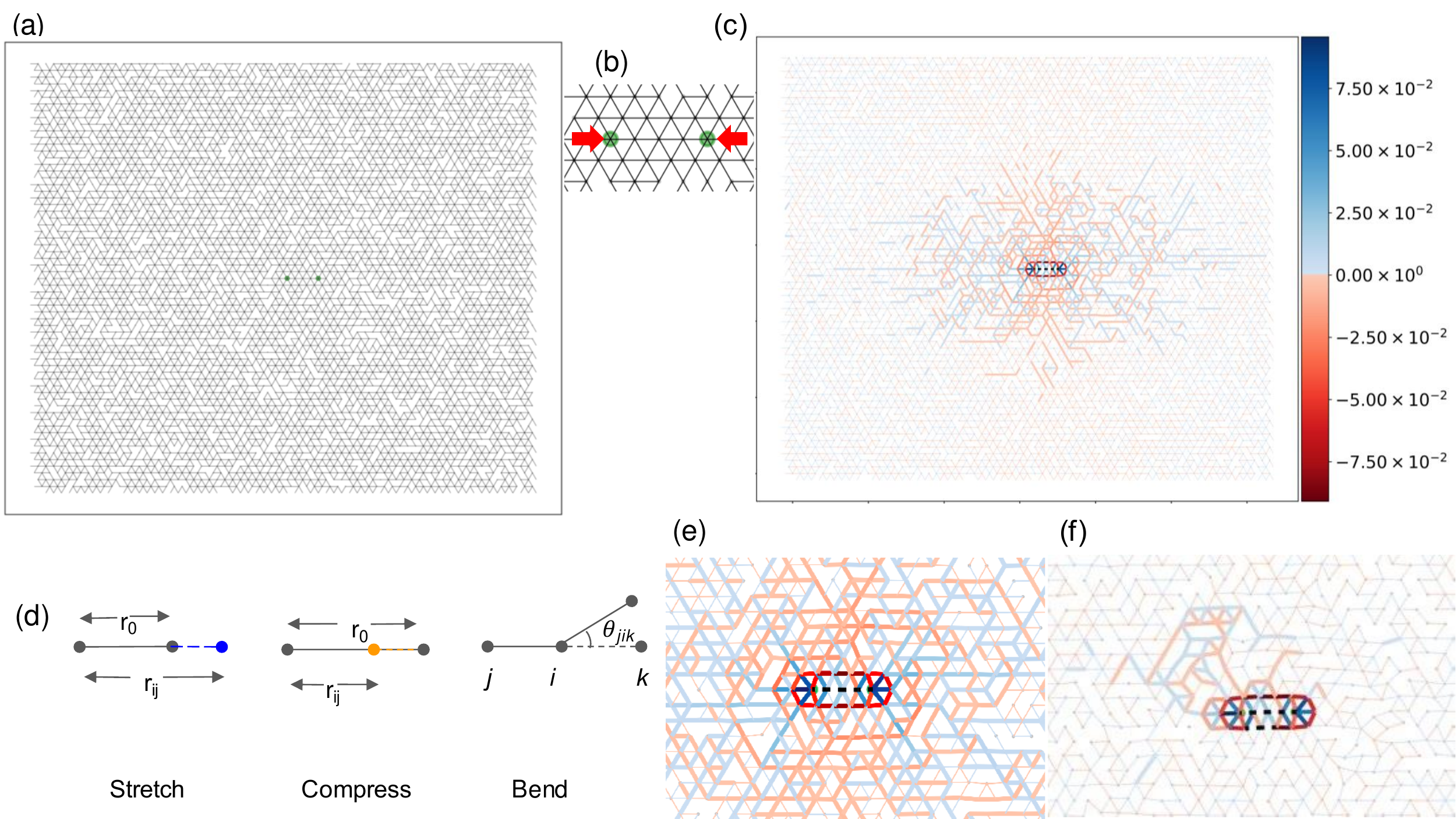}
            \caption{ \textbf{Elastic fiber network model and representative minimum energy configurations from simulation.}
            (a) An example hexagonal lattice (comprising equilateral triangles) of size $64 \times 64$ with bonds randomly removed according to the bond occupation probability, $p = 0.8$, is shown. The two nodes marked in green are those on which we apply a pair of equal and opposite forces, $f$, here along the $x-$axis, to create a force dipole. (b) Zoomed in version of (a) with red arrows indicating the direction of the active dipole forces. (c) A sample lattice simulation with $p$ = 0.8 deformed by a single dipole, with value of force, $f$ = 0.4. The colorbar shows strain values for stretched (blue) and compressed (red) bonds. Bonds that carry a strain magnitude above a threshold value
            $\epsilon_{0} =0.001$ are highlighted in bold. The bonds with relatively very high/low strains, usually found very close to the dipole, are plotted with black dashed lines to make sure that there is appreciable color gradient visible in the rest of the bonds. (d) Schematic illustrating the stretching, compression and bending of bonds that contribute to mechanical energy in the model. Each bond, when present between neighboring nodes, is modeled as a linear (Hookean) spring with a uniform rest length and stiffness, $\mu$. Each pair of collinear bonds is associated with an angular spring of stiffness $\kappa$ that penalizes deviations in angle between the two bonds from the original value, and represents fiber bending stiffness. Buckling and other forms of elastic nonlinearity of bonds are ignored, which is justified for small forces and stiff fibers. (e) Zoomed in view of (c), showing bonds with strains higher than the threshold in thick red (compressed) or blue (tensile). The bonds are shown as thick lines if the strain is above 
            $0.001$. (f) Same as (e) but for $p = 0.6$. Local bending of bonds is visible that is not visible in (e) }
            \label{fig:Model}
        \end{figure*}

We model the elastic medium as a regular hexagonal lattice (of equilateral triangles) from which bonds can be randomly removed to create disordered fiber networks. 
Such a lattice  (Fig.~\ref{fig:Model}) models an actin fiber network where a set of collinear bonds is understood to represent a fiber. Each bond between a pair of nodes can stretch or compress with respect to its initial rest configuration. Fiber bending is implemented through the rotation or relative change of angle between two collinear bonds. Individual bonds in this model do not bend or change shape.  The actin network is deformed by myosin motor activity.   Each actomyosin unit, comprising a myosin motor aggregate and the actin filaments it binds to, is modeled as a contractile force dipole: a pair of equal and opposite forces applied at two nearby nodes. We create a force dipole by applying equal and opposing forces at two selected nodes in the network (\ref{fig:Model}b). In depleted networks, we also make sure that bonds in the vicinity of the force dipole are not removed, to allow efficient force transmission and to prevent large node displacements due to local floppy modes.  The nodes on which we apply forces are shown as green circles and the direction of forces are shown by red arrows. This application of a force dipole mimics how myosin motors contract the actin cytoskeleton and deform it \cite{wang2012active, broedersz_11}. 

The network deformation is calculated by minimizing its total mechanical energy, which includes the elastic energy stored in the network from the stretching, compression and bending of bonds \cite{das2012redundancy}, as well as the work done by the force dipole in moving pairs of corresponding nodes towards each other \cite{Ronceray16} 
\begin{equation} \label{eq:total energy_main}
\begin{split}
E_t &=  E_{s} + E_{b} + E_{d} = \frac{\mu}{2}\sum_{\langle ij \rangle}  (r_{ij}-r_{0})^2 +  \\
&\quad    \frac{\kappa}{r_{0}} \sum_{\langle jik \rangle} {2\sin^2(\theta_{jik}/2)} + \sum_{\langle mn \rangle} \mathbf{F} \cdot \mathbf{l}_{mn}
\end{split}
\end{equation}

where $\mu$ and $\kappa$ are the stretching and bending moduli respectively, $r_{ij}$ represents the length of the bond connecting two neighboring nodes, and $r_{0}$ is its rest length, set to unity for all bonds in our system.  Fiber bending is modeled by angular springs at every pair of collinear bonds, which are defined by the three nodes, $j$, $i$ and $k$, with $i^{th}$ node being central. For each such pair of collinear bonds, we impose an energy cost, 
$E_{b,jik} = (2\kappa/r_{0})\sin^2(\theta_{jik}/2)$
for deviations from collinearity, which is the discretized form of the bending energy in the worm-like-chain  model for semiflexible polymers.  The dipole energy is the scalar product of force applied  on and the separation vector between the pair of nodes comprising the dipole, which need not in general be along a bond.  Here, $\mathbf{l}_{mn}$ is the  separation vector connecting the $m^{th}$ and $n^{th}$ nodes that comprise one force dipole. The force is central and always along the separation vector $\mathbf{l}_{mn}$ , though myosin motors may also act transversely to fibers \cite{ronceray2019fiber}.  In this work, unless otherwise stated, the dipole nodes are chosen to be $4$ nodes apart, \emph{i.e.} $l_{mn} = 4 r_{0}$.

\begin{table*}
\centering
\begin{center}
\begin{tabular}{ |c|c|c|c| } 
 \hline
 Parameter & Symbol & Simulation Value &  Physical Value \\ 
 \hline
Fiber length  & $l$ & $1$ & $10 \mu m, 1 \mu m, 100 nm $\\ 
 Bending:stretching rigidity & $\widetilde{\kappa}$ & $10^{-6}, 10^{-4}, 10^{-2}$  & $ 17 \, \mu m \cdot k_B T $\\ 
 Myosin force & $f$ & $0.4$  & $ 0.4 \,$ pN\\ 
 \hline
\end{tabular}
\caption{\textbf{Model Parameters} By changing the bending-to-stretching stiffness ratio, we represent three different typical actin fiber lengths \cite{phillips,green_12}. These ranges of bending-stretching ratios have been used in previous works \cite{das2012redundancy, popov2016medyan}.
The parameters listed here allow us to investigate rigid as well as floppy networks. The dipole force is comparable to that produced by a nonmuscle myosin motor minifilaments \cite{howard}. }
\end{center}
\label{table:parameters}
\end{table*}

Our 2D triangular network of nodes connected by bonds has periodic boundary conditions along both $x$ and $y$ directions. The lattice size used for the simulations  is $64 \times 64$, unless stated otherwise. Force is applied incrementally to the nodes of the dipole. At each step of force application, we use the conjugate gradient method to minimize the energy of the network given in Eq. ~\ref{eq:total energy_main} and find its new configuration \cite{PressW}. 
The energy tolerance to accept the new configuration as being the energy minimum  is set to $2 \times 10^{-6}$. The process is repeated up to a maximum force value and for a fixed number of force iterations. For the results reported here, forces were applied at increments of $f_{0} = 0.04$ to each dipole node, for a total of $10$ steps. The resulting force applied is then $f = 0.4$. 




For an elastic fiber, the bending modulus ${\kappa} = \frac{\pi}{4} E a^{4}$ depends on the Young's modulus $E$ and  the fiber radius, $a$ \cite{landau_lifshitz_elasticity}.
The stretching modulus is $\mu = \pi E a^{2}$.  We define a nondimensional ratio of bending to stretching, which depends on the fiber length: $\widetilde{{\kappa}} = {\kappa}/(\mu l^{2}) = \frac{1}{4} a^{2}/l^{2}$. For an actin filament that has a diameter of $7$ nm \cite{phillips}, and is of length $l \sim 10$ $\mu$m, we estimate $\widetilde{{\kappa}} \sim (a/l)^2 \sim 10^{-6}$. For shorter filaments with lengths $l \sim 1 \mu$m or $l \sim 100$ nm, such as those found in the actin cortex \cite{green_12}, this value is larger, $\widetilde{{\kappa}} \sim 10^{-4}$ and $\widetilde{{\kappa}} \sim 10^{-2}$, respectively.
This range of values is consistent with bending to stretching ratios previously used for modeling actin filaments \cite{das2012redundancy, popov2016medyan}.

Using the worm-like chain model for semiflexible polymers, we estimate a thermal fluctuation induced strain of $\epsilon_{0} = l/(6 l_p)$ \cite{broedersz_14}. For an actin filament of length $l \sim 1$ $\mu$m and known persistence length $l_{p} \sim 17$ $\mu$m, this strain has value of $\epsilon_{0} \sim 10^{-2}$, while for $l \sim 100$ nm, $\epsilon_{0} \sim 10^{-3}$. We use this as a threshold strain to identify the range of force transmission from the source dipole in our model fiber networks. The elastic force corresponding to this strain is $\mu \cdot \epsilon_{0} \sim \kappa l/(6 l_{p} a^{2}) \sim k_{B} T/(6a) \cdot (l/a)$, which for $l \sim 100$ nm, $a = 7$ nm and $k_{B} T \approx 4\, pN\cdot nm$ is the thermal energy scale at room temperature, gives a value of the characteristic value of force $\sim$ pN.  This is comparable with the force produced by a nonmuscle myosin motor minifilament with $~10$ heads, each producing a force of $~$pN, with a duty ratio of $\sim 10 \, \%$ \cite{howard}. Thus, a force of $f =0.4$ applied in the simulation corresponds to $\sim 0.4$ pN, while the separation of $4$ units corresponds to $4 l \sim 400$ nm, which is indeed the typical size scale for an actomyosin contractile unit in stress fibers of the cellular cytoskeleton \cite{hu_17}.



\section{Results}

\subsection{Single dipoles in over-coordinated networks}
\begin{figure*}[htp]
            \centering
            \includegraphics[width=16cm]{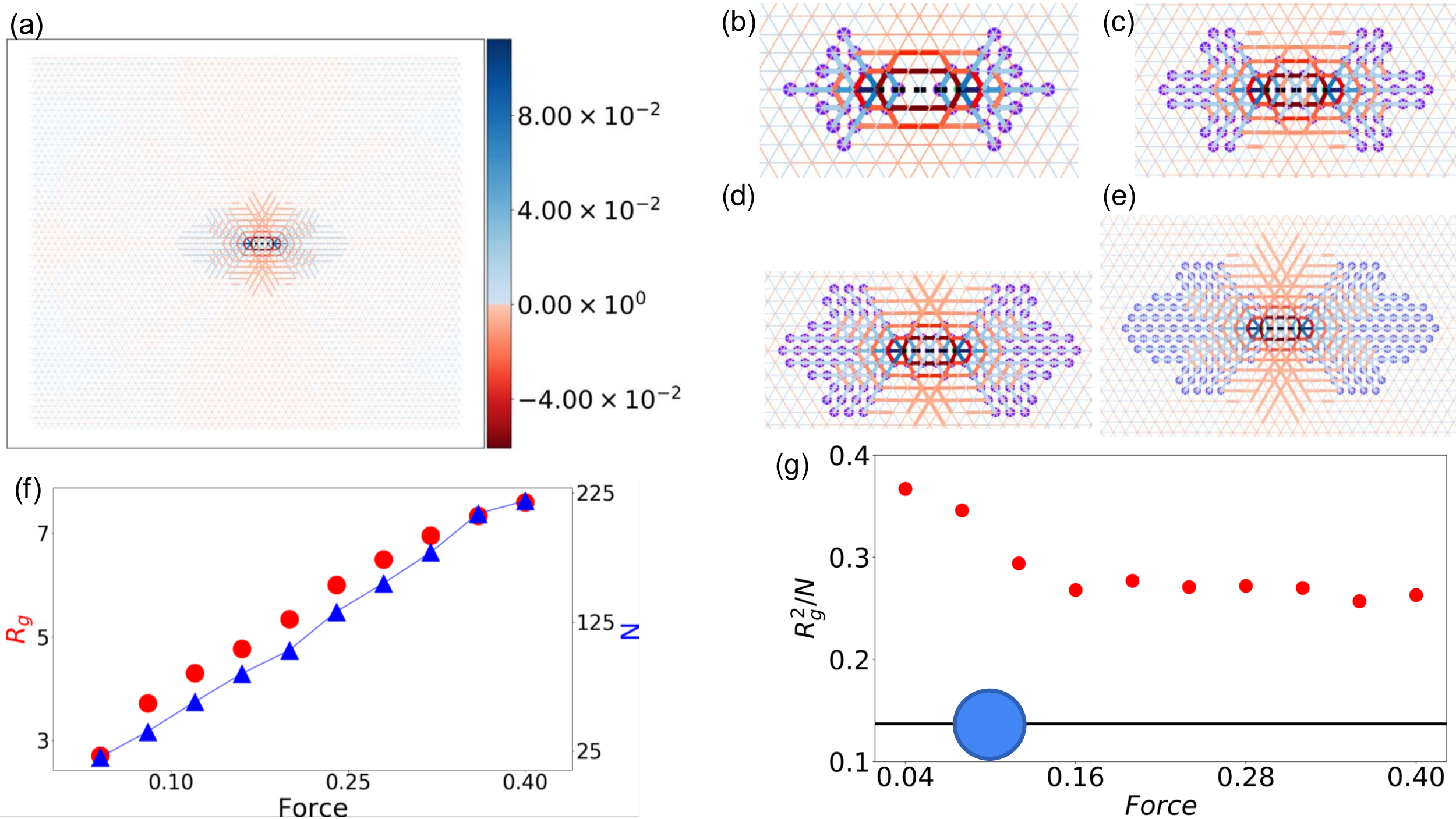}
            \caption{ \textbf{Strain clusters as measures of force transmission from a single dipole in a regular lattice.} (a) Representative network configuration and bond strains induced by a single dipole in an undiluted, regular triangular lattice ($p=1$) for $f = 0.4$. The color bar shows the strain in the bonds, where blue (red) represents compression (tension). Thick bonds represent  highly strained bonds that carry strain above the threshold value of $ \epsilon_{{th}}$ $= 0.003$. (b,c,d,e) Cluster of nodes (blue) connected to strongly tensile bonds (above strain threshold) for increasing values of dipole force: $f = 0.08, 0.16, 0.24$ and $0.4$. (f) The size of each tensile cluster is characterized using two metrics: the number of nodes ($N$) in the cluster (blue data points), as well as its radius of gyration (red data points), $R_{g}$ (eq. ~\ref{eq:radius of gyration}). Both measures show that the cluster size grows with increasing value of the dipole force. (g) Ratio of radius of gyration squared to number of nodes in tensile cluster, which is a measure of the deviation from circularity (or the anisotropy) of the cluster shape, as a function of the applied dipole force. For reference, the horizontal line is the expected value for a circular region in a hexagonal lattice.}
            \label{fig:Single dipole}
        \end{figure*}

We first consider the simplest case of a single force dipole in a uniform triangular lattice of springs. The effective bending stiffness set to be very small in relation to stretching, $\widetilde{\kappa} = 10^{-6}$.
The resulting network deformation  is shown in (Fig.~\ref{fig:Single dipole} a) with stretched (compressed) bonds colored in blue (red). Additionally, strongly strained bonds beyond a chosen threshold are highlighted in bold.  As seen here, a force dipole creates primarily stretched regions along its axis, \emph{i.e.} to its left and right, in this case. Similarly, the dipole compresses the network in the transverse direction, in this case, above and below it. 

To quantify the spatial extent of force propagation in the network from the dipole, we analyze clusters of nodes that are connected to strongly strained bonds. A cluster is defined to be a set of nodes, with positions ${{\bf{r}_{i}}}$, that share at least one bond with magnitude of strain above a threshold value. For the results presented here, the threshold value is set to $\epsilon_{0} = 0.003$. This is comparable to our estimate for strain induced in a semiflexible actin polymer by thermal fluctuations alone.  We verified that the results do not qualitatively change for a smaller strain threshold (Fig. S1). 
The cluster nodes may or may not be directly connected to the dipole nodes, though in practice we observe that most highly strained bonds form a single, large, connected cluster that includes the dipole nodes. The radius of gyration for a cluster with $N$ nodes is defined below in the standard way,
\begin{equation} \label{eq:radius of gyration}
R_{g} = \sqrt{\frac{1}{N} \sum_{i=1}^{N} {|\bf{r}_{i} - \bf{r}_{c}|}^{2}}
\end{equation}
where $ \bf{r_{i}}$ is the position of the $i^{th}$ node in the cluster, and $\bf{r_{c}} = \frac{1}{N} \sum_{i=1}^{N} \bf{r}_{i} $ is the center of mass of the cluster.

We first examine the dependence of the network deformation on dipole force.
Fig.~\ref{fig:Single dipole} (b,c,d,e) show network configurations for a single dipole at representative values of force corresponding to $f=0.08, 0.16, 0.24$ and $0.4$, respectively. The nodes adjacent to bonds that are strongly stretched ($\epsilon > \epsilon_{0} =0.003$) are colored in blue and comprise a single tensile cluster around the dipole. Both the network configurations in Fig.~\ref{fig:Single dipole} (b,c,d,e) and the quantification of cluster size in Fig.~\ref{fig:Single dipole}f in terms of the number of nodes in the tensile cluster, $N$, and radius of gyration, $R_{g}$, show a linear increase in cluster size with dipole force, $f$.  This is expected since the strain in a linear elastic medium induced by a force dipole is proportional to the dipole stress, and thus its force density.

 To quantify the trends in cluster shape, we introduce a parameter $R_{g}^{2}/(N r_{0}^{2})$ that measures the deviation of cluster shape from a compact circle. Higher values of this parameter correspond to more anisotropic or branched shapes.
This value shows that the cluster becomes slightly less elongated as the force increases (Fig.~\ref{fig:Single dipole} and saturates at a force of about $0.26$, well above the limiting value of  for a circular region in a 2D triangular network, given by $ \sqrt{3}/{4\pi} = 0.137$ (Fig. S2) and marked by the horizontal line in Fig.~\ref{fig:Single dipole}g. 

 \begin{figure*}[htp]
            \centering
            \includegraphics[width=16cm]{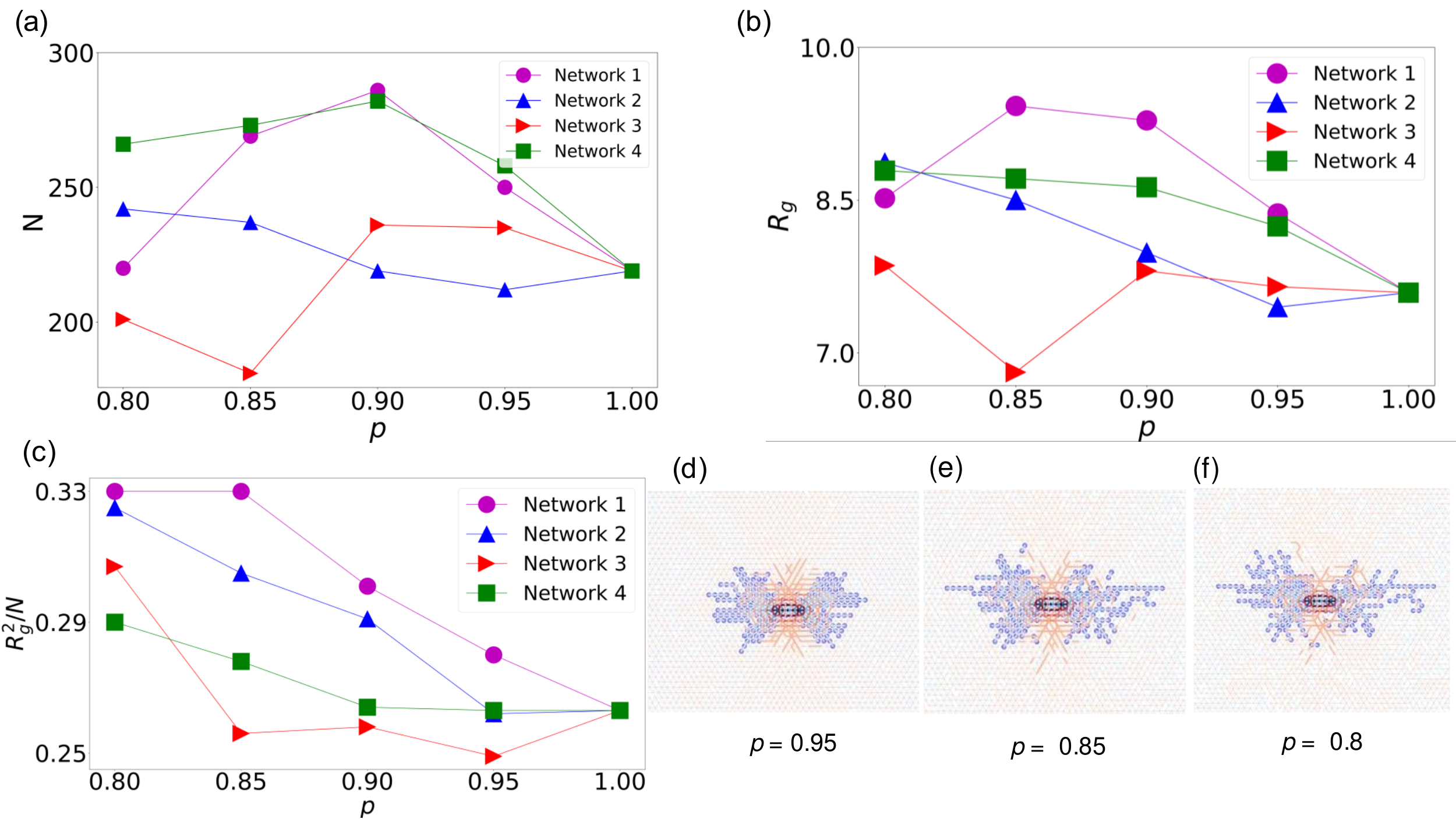}
            \caption{ \textbf{Tensile cluster around single dipole in slightly diluted, stretching-dominated networks. } (a-c) Tensile strain cluster size trends for four different networks on successive bond dilution. (a) Number of nodes in the cluster ($N$), (b) Radius of gyration of cluster ($R_{g}$), and (c) $R_{g}^{2}/N$, a measure of compactness of cluster shape, all as functions of the bond occupation probability $p$ in the stretching dominated ($ p > p_{cf} = 0.67$) regime. While cluster size is apparently insensitive to network connectivity in the stretching dominated regime, the cluster shape becomes less circular as network connectivity is reduced. (d,e,f) 
            Representative configurations of tensile clusters at $p = 0.95, \; 0.85$ and  $0.8$ for Network $4$ confirm this trend and show that clusters elongate along the dipole or $x-$axis as more bonds are removed.}
            \label{fig:Single dipole cluster}
        \end{figure*}

To examine how the force propagation in the network is affected by a small amount of disorder, we dilute the network by randomly removing bonds, while staying in the stretching-dominated regime ($p \geq 0.67$).  Since this procedure may result in many different specific network realizations, we simulate four different networks at each value of $p$ to show the fluctuation in strain and cluster trends. In each case, we ensure that we do not have any singly-connected nodes, to prevent dangling bonds, and also ensure that there are no unphysical metastable states when minimizing the network energy.  This is done by changing the force value by different step sizes, which we show all result in the same values of final energy (see Fig. S3).

 Similar to Fig.~\ref{fig:Single dipole} for the $p=1$ uniform network, we consider the tensile force clusters for these slightly diluted networks in Fig.~\ref{fig:Single dipole cluster}. We notice that the cluster size does not show any significant change as the network is diluted, though there is possibly a slight increase in the average $R_{g}$ (Fig.~\ref{fig:Single dipole cluster} a,b). This suggests that  while there are fewer bonds near the dipole as $p$ decreases, there are new bonds, that previously did not participate in the tensile cluster,that now become highly strained. Altogether, these two effects keep the size of the cluster roughly constant. However, the cluster shape clearly becomes more anisotropic as bonds are diluted (decreasing $p$), as shown by Fig.~\ref{fig:Single dipole cluster} c). 
This is also apparent in the simulated network configurations shown for three different bond dilution factors (Fig.~\ref{fig:Single dipole cluster} d, e, f). As we deplete the uniform network, there is an increase in the number of nodes participating in the tensile cluster along the axis of the dipole that manifest as branching ``force chain'' structures.
This increase in cluster extent along the axis of the dipole can be attributed to the fact that most of the stretched bonds (blue) lie to the left and right of the force dipole (Fig.~\ref{fig:Single dipole} e). At the same time, because some bonds are randomly removed, nodes that made up the tensile cluster in a uniform network may not be a part of the cluster anymore. 
While we do not track each force chain individually as in Ref.~\cite{mann2019force}, our clustering analysis gives a rough measure of the force chain length in the form of the $R_{g}$.  Altogether these results suggest, that for stretching-dominated networks, the range of tensile strains increases along the dipole axis, with bond dilution.  

        \begin{figure*}[htp]
            \centering
            \includegraphics[width=16cm]{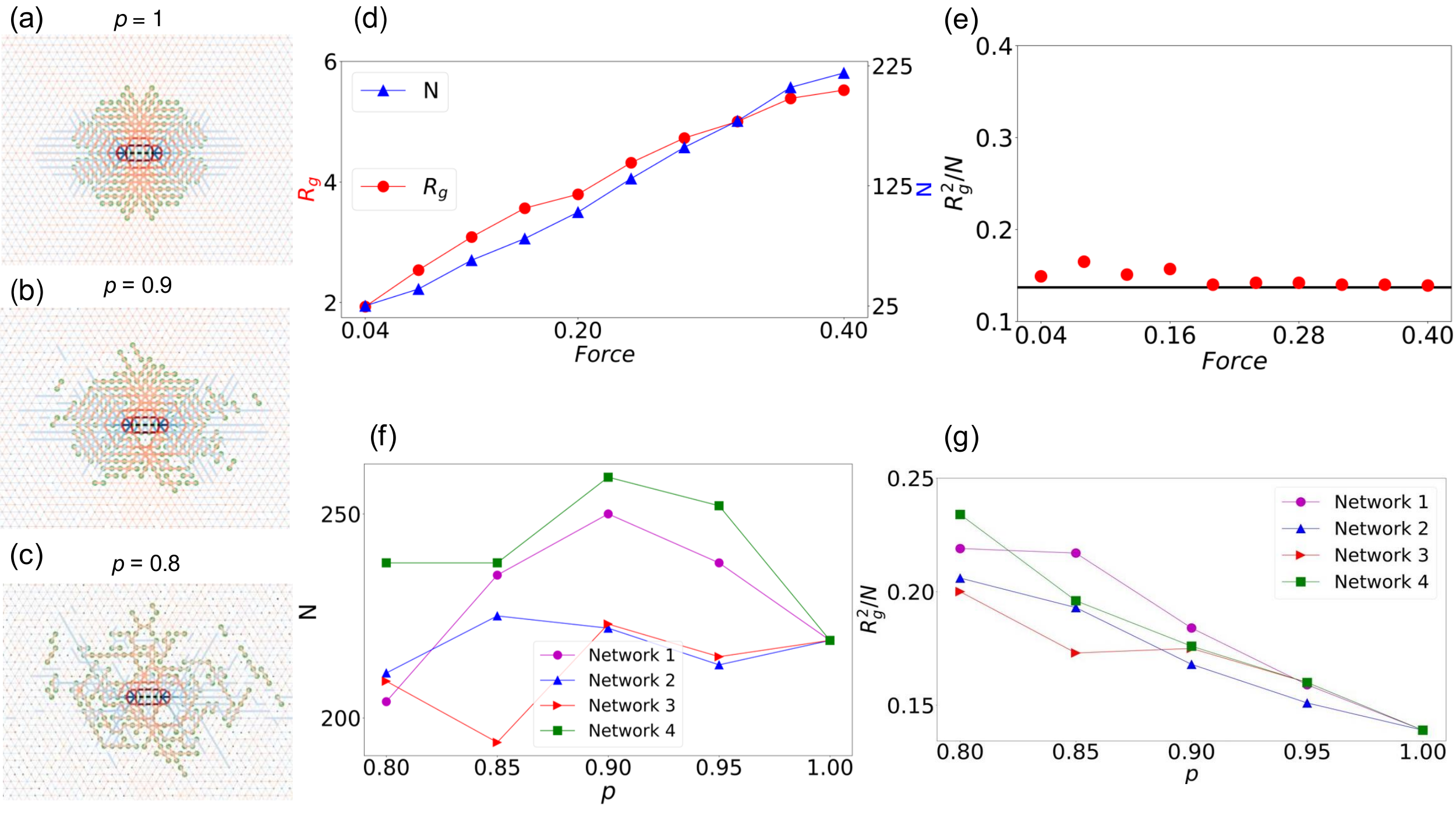}
            \caption{ \textbf{Compressive strain cluster around a single dipole in slightly diluted, stretching-dominated networks. } (a-c) Clusters of nodes (green) connected to highly compressed (strain below prescribed threshold, $\epsilon < -0.003$) in typical network configurations. (a) Cluster for $p = 1$ is approximately circular. (b,c) Clusters deviate from a compact circle to having ring like and disconnected structures upon bond dilution. (d) Number of nodes in the cluster ($N$) and radius of gyration of cluster ($R_{g}$) for $p = 1$ confirm increasing cluster size with force. (e)
            $R_{g}^{2}/N$, a measure of cluster shape, approaches $0.137$ for $p = 1$ (shown by horizontal grey line), the value for a circular shape. (f) Number of nodes stays relatively same in all four networks as $p$ changes. (g) However, the cluster shape deviates from circularity as we deplete the network.}
            \label{fig:Single dipole cluster compressive stretch regime}
        \end{figure*}

Since the force dipoles induce anisotropic elastic deformations, we expect different spatial distribution of compressed and stretched bonds. To characterize this difference,  we now consider compressive clusters that comprise nodes connected to bonds with larger compressive (negative) strain, $\epsilon < -\epsilon_{0}$, in Fig.~\ref{fig:Single dipole cluster compressive stretch regime}. For the uniform ($p=1$) network shown in Fig.~\ref{fig:Single dipole cluster compressive stretch regime}a, the cluster is almost circular in shape. This structure is different from tensile clusters (Fig.~\ref{fig:Single dipole}) which are anisotropic with major axis aligned along the dipole direction. It is also notable that island clusters and ring like structures are formed when the network is depleted. These are seen for the $p=0.9$ case shown in Fig.~\ref{fig:Single dipole cluster compressive stretch regime}b, and become particularly pronounced at $p=0.8$ in  Fig.~\ref{fig:Single dipole cluster compressive stretch regime}c.
These ring like structures are a signature of compressive clusters in our simulations, while the tensile clusters have linear force chain like structures (Fig.~\ref{fig:Single dipole cluster}). We further verify in Fig.~\ref{fig:Single dipole cluster compressive stretch regime} that the size of a compressive cluster increases linearly with increasing force as expected. The shape parameter, $R_{g}^{2}/N$, for the compressive cluster in a $p=1$ network is unaffected by force and remains very close to the circular value (horizontal line), as shown in Fig.~\ref{fig:Single dipole cluster compressive stretch regime} e.   

Similar to tensile clusters, the removal of bonds leads to decrease in number of cluster nodes, while new compressive bonds with strains below $- \epsilon_{0}$ participate in the cluster. Together, these two effects balance and result in no net systematic change in cluster size with small amounts of depletion, as shown in Fig.~\ref{fig:Single dipole cluster compressive stretch regime}f. Lastly, we show in Fig.~\ref{fig:Single dipole cluster compressive stretch regime}g that depletion increases the shape anisotropy, $R_{g}^{2}/N$, corresponding to the clusters becoming less circular. This deviation from circularity is due to the voids as well as islands forming for the compressive clusters, as seen in Fig.~\ref{fig:Single dipole cluster compressive stretch regime} c). We also note that the shape parameter is appreciably lower for the compressive clusters ($R_{g}^{2}/N \approx 0.15-0.25$) than for the tensile clusters ($R_{g}^{2}/N \approx 0.25-0.33$) in Fig.~\ref{fig:Single dipole cluster}.  Altogether, we show that the propagation of compressive strains from the dipole is qualitatively different from that of tensile strains. This may affect the way a second, test dipole interacts with the first dipole. Since a contractile dipole lowers the network deformation when it is in a stretched region, these strain maps may guide the favorable position of a second dipole with respect to the first. However, our results suggest that such favorable configurations of two dipoles may be sensitive to the specific network.

\subsection{Single dipole in under-coordinated networks}

        \begin{figure*}[htp]
            \centering
            \includegraphics[width=16cm]{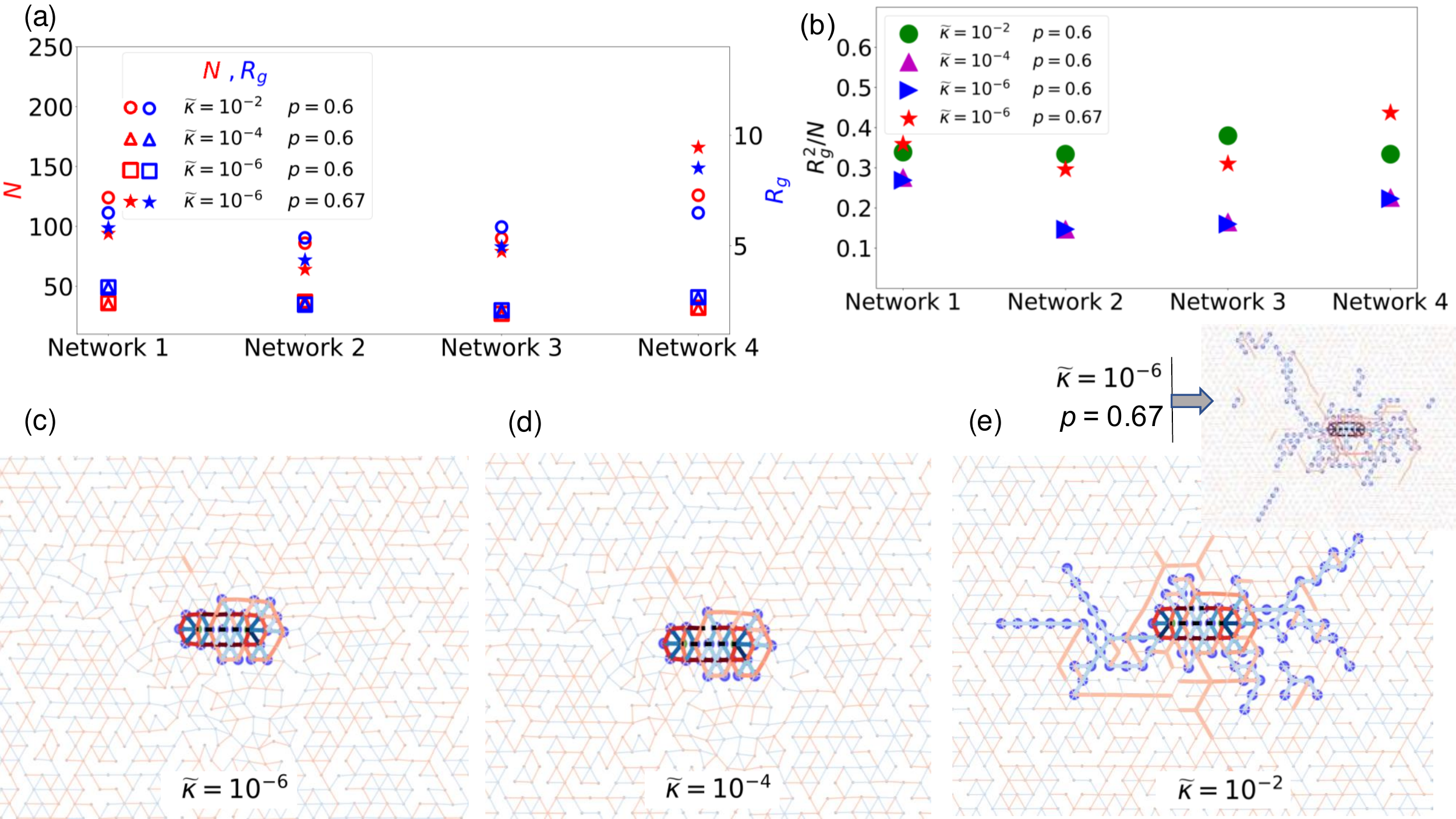}
            \caption{ \textbf{Strain cluster around single dipole in highly diluted, bending-dominated networks} All plots are for four network realizations, diluted below the isostatic limit. The network deformation is dominated by low energy (floppy) modes involving bonds bending (rotating) in preference to stretching/compressing. 
            (a) 
            Number of nodes (red data) and radius of gyration (blue data) in each tensile cluster  show two distinct regimes: for $\widetilde{\kappa} = 10^{-6}$ and $10^{-4}$, the clusters are very small compared to the high bending stiffness $\widetilde{\kappa} = 10^{-2}$, case. Low bending stiffness case ($\widetilde{\kappa} =10^{-6}$)  at the bend-stretch transition point, $p=0.67$) behaves like higher bending stiffness case. (b) Cluster shape parameter ${R_{g}}^{2}/N$ also shows a gap between the higher bending stiffness, $\widetilde{\kappa} = 10^{-2}$,and lower bending stiffness cases. Low bending stiffness case ($\widetilde{\kappa} =10^{-6}$)  at the bend-stretch transition point, $p=0.67$) behaves like higher bending stiffness case. (c,d,e) Tensile clusters in the same network at different bending stiffness values, $\widetilde{\kappa} = 10^{-6}, 10^{-4}$ and  $10^{-2}$. The network at $\widetilde{\kappa} = 10^{-2}$ allows tensile/compressive strains to be transmitted farther than for networks in the pure bending regime. This response is similar to the bend-stretch coupled regime accessed by keeping $\widetilde{\kappa} = 10^{-6}$ and increasing $p$ to $p_{cf} = 0.67$ (inset e). 
            }
            \label{fig:Single dipole bend}
        \end{figure*}

Upon further dilution of bonds, specifically for $p<p_{cf} \approx 0.67$, the network enters an under-coordinated regime. There are many available low-energy bending modes for such networks, which allow nodes to move in response to the dipole forces such that collinear bonds bend, but stretching (or compression) of bonds is minimal.    In this bending-dominated regime, at $p = 0.6$,  
, we see in Fig.~\ref{fig:Single dipole bend}a that the tensile clusters are small ($N \sim 50$ here compared with $N\sim 200$ in the stretching-dominated regime) in all of the networks at $\widetilde{\kappa} = 10^{-6}$. Increasing the bending stiffness relative to stretching to $\widetilde{\kappa} = 10^{-4}$ does not have a pronounced effect, either on cluster size or on the shape parameter, as seen in Figs. ~\ref{fig:Single dipole bend}a-b. These cluster trends are easily seen in the sample network configurations shown in  Figs. ~\ref{fig:Single dipole bend}c-d.  Significantly compressed (red) or stretched (blue) bonds occur only in the immediate vicinity of the force dipole. Some of these bonds (deep red and blue) carry higher strain because the dipole nodes are locally over-coordinated. 
Away from the dipole, the bonds are not strained but the network shows significant bending deformations. This is because local clusters of bonds can easily rotate to 
reduce strains, especially in regions of lower local connectivity. 

However, a qualitatively and quantitatively different behavior is seen, when the bending to stretching stiffness ratio increases to  $\widetilde{\kappa} = 10^{-2}$. Figs. ~\ref{fig:Single dipole bend}a-b show that both cluster shape and anisotropy are significantly enhanced at this value of $\widetilde{\kappa}$.  Compressive clusters show a similar behaviour (see Fig S4).  This is visually confirmed by the sample simulated network configuration shown in Fig.~\ref{fig:Single dipole bend}e, which resembles the tensile cluster seen for a single dipole in the stretching-dominated regime. Both tensile and compressive force chains are clearly seen to extend from the dipole nodes. This suggests, that due to the higher energy cost of bending, the bonds are not as free to rotate and relax stretching as for the lower $\widetilde{\kappa}$ networks.  In fact, previous simulations show the existence of such a bend-stretch coupled regime, where both bending and stretching deformations occur in response to network shear \cite{broedersz_14}. The network shear modulus in this intermediate regime scales with both $\kappa$ and $\mu$.  These works showed that the bend-stretch coupled regime occurs in the transition between the bending (low $p$) and stretching (high $p$) dominated regimes 
, and that the range of $p$-values over which this regime occurs grows wider as the bending to stretching ratio, $\widetilde{\kappa}$, is increased \cite{broedersz_14} .  Motivated by the prediction for the existence of the bend-stretch regime close to the transition point, we simulate the network deformation at $p =0.67 \approx p_{cf}$ for the lower $\widetilde{\kappa} = 10^{-6}$. Indeed, we see that for this case (inset to Fig.~\ref{fig:Single dipole bend}e), the bend-stretch coupled behavior characterized by moderately large cluster size is restored. The measurement of cluster shape (Fig. ~\ref{fig:Single dipole bend}b) also shows a quantitative agreement between networks that are stiff to bend ($\widetilde{\kappa} = 10^{-2}$) at $p=0.6$, and  and networks that are softer to bend ($\widetilde{\kappa} = 10^{-6}$) but are closer to the bending-stretching transition, $p=0.67$.
Thus, the bend-stretch regime occurs either when network connectivity approaches $p_{cf}$, or when it remains in the under-coordinated regime $p<p_{cf}$ but has higher bending stiffness $\widetilde{\kappa}$. Our simulation results  show that large compressive and tensile force clusters emerge in the bend-stretch coupled regime, as seen also in simulations of networks under shear where force clusters get bigger as the transition $p =p_{cf}$ is approached. These clusters restore long-range force transmission through the network, making them comparable with the stretching-dominated cases.



\subsection{Strain distribution for single dipoles}

\begin{figure*}[htp]
            \centering
            \includegraphics[width=16cm]{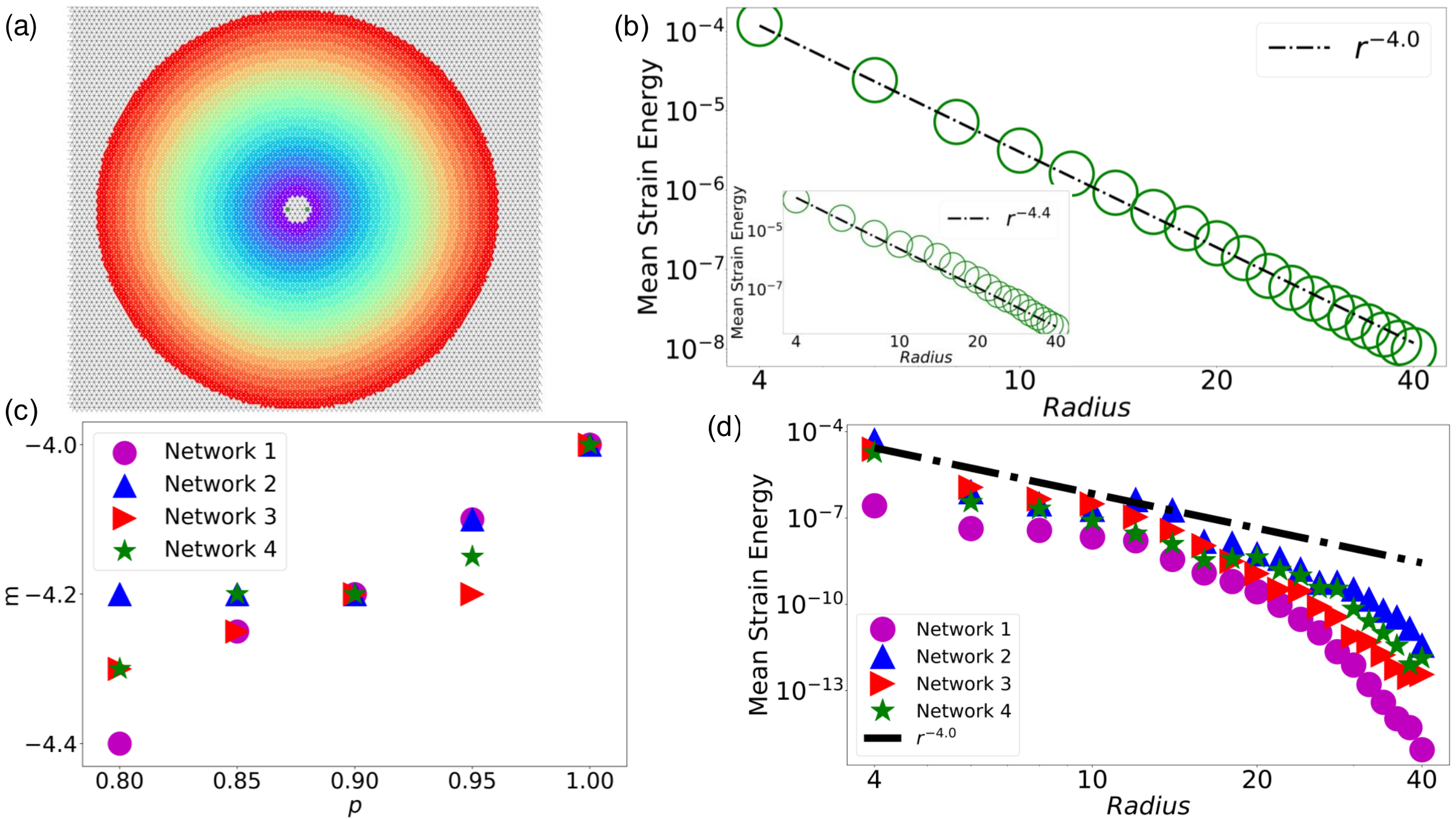}
            \caption{ \textbf{Decay of strain energy in network bonds with distance from a single dipole } 
            (a) The strain energy from stretching/compression of bonds is calculated over annular regions of width $\Delta R =2$, shown in different colors in the undeformed, uniform network. To increase sampled distances from the force dipole (center), we consider simulation box of increased size $L = 96$. (b)  Mean strain energy, $E_{strain}^{k} =   \mu/2  \langle  \epsilon_{j}^2 \rangle $, averaged over each $j^{th}$ bond in the $k^{th}$ annular region, vs radius of the annulus, $R_k$, for a uniform network.  Green circles are measured from simulation data, while the dot-dash line shows for comparison, a decay rate in strain energy of $r^{-4}$, predicted from linear elasticity theory. Inset: strain energy decay with radial distance in Network 1 for disordered network with $p = 0.8$.
            (c) For depleted networks in the stretching regime, we plot the exponents of the best fit power law for decay of strain energy vs distance, for all four network realizations, against the corresponding $p$ values. The value of 
            slopes are close to the value of $-4$ for uniform networks, decreasing slightly with bond dilution.
             (d) Mean strain energy in each annulus vs. radius of annular region, for networks in the bending-dominated regime ($p = 0.6$). There is no single power law regime in the decay of strain energy, which shows slow decay close to the dipole, and then more rapid decay at larger distances. The strain energies are also lower in value when compared to networks in the stretching regime (see b), showing that bending screens out strain propagation making it shorter-ranged. 
             }
            \label{fig:Single dipole strain energy}
        \end{figure*}
        
In addition to force clusters, we may quantify the range of strain propagation in the network in terms of the rate of decay of elastic strain energy with distance from the dipole. In order to measure the average strain energy density at a given radial distance from the dipole, we
consider annular regions of increasing radii ranging from $R=4-40$, each of thickness $\Delta R =2$, and centered midway between the nodes of the single dipole, as shown in Fig.~\ref{fig:Single dipole strain energy}a. The maximum radius of the annular region is set by the total lattice size, which was chosen to be $L=96$ for this particular measurement, in order to allow a wider range of distances.
We then calculate the average strain energy in all bonds in the $k^{th}$ ring as $E_{strain}^{k} =  \mu/2  \langle \ \epsilon_{ij}^{2} \rangle$, where $i,j$ represent all adjacent nodes connected by bonds in the $k^{th}$ ring, and $\epsilon_{ij} = (r_{ij}/ r_{0} -1)$ is the corresponding bond strain.   $E_{strain}^k$ is thus the mean strain energy stored in the $k^{th}$ annular region. 

From this analysis, we find that the mean strain energy decays as a $r^{-4}$ power law  with distance for a uniform network as seem from Fig.~\ref{fig:Single dipole strain energy}b. This is expected from linear elasticity theory because the strain energy density $E_{strain} \propto \epsilon^{2}$, where the continuum strain field induced by a force dipole in an infinite, 2D elastic medium decays with distance as $\epsilon \propto \frac{1}{r^{2}}$. Taken together, this predicts $E_{strain} \sim r^{-4}$. See Appendix B for the continuum linear elasticity derivation of the full strain tensor for a force dipole. 
We find that on introducing small amounts of disorder ($p<1$), the strain energy decay begins to deviate from this $r^{-4}$ scaling.  For example, the inset of Fig.~\ref{fig:Single dipole strain energy}b shows the strain energy decay for $p=0.8$ for a specific network realization (Network 1).
In general, we find that the strain energy decay remains a power law of the form $r^{-m}$, with the decay exponent $m$ remaining close to that of the uniform $p=1$ network in stretching-dominated regime as shown in Fig.~\ref{fig:Single dipole strain energy}c. 
For stretching-dominated networks, the rate of decay of strains increases with increasing depletion, as seen from the power law decay exponents in Fig. ~\ref{fig:Single dipole strain energy}c. The relatively higher localisation of strains in the vicinity of the dipole for more depleted networks leads to a higher rate of decay in the strains away from the dipole.


While the stretching-dominated networks all show power law decays of strain with distance, bending dominated networks behave qualitatively differently, as shown in from Fig.~\ref{fig:Single dipole strain energy}d  for the four networks with $p=0.6$ and $\widetilde{\kappa} = 10^{-6}$. In these networks, the strain decays faster at larger distances compared to smaller distances and there is no single power law regime. On comparing with the representative strain map for this case shown in Fig.~\ref{fig:Single dipole bend}c, we suspect that this may be because while bonds very close to the dipole can be strongly stretched or compressed, the long-range response is dominated by bending of bonds. In this region, the mean strain energy of the bonds decays rapidly, as the bonds are barely stretched.  The decay of elastic energy with distance is thus very different for the bending-dominated network from a continuum, linear, elastic theory prediction, which underscores the non-affine nature of deformations in this regime.

        \begin{figure*}[htp]
            \centering
            \includegraphics[width=16cm]{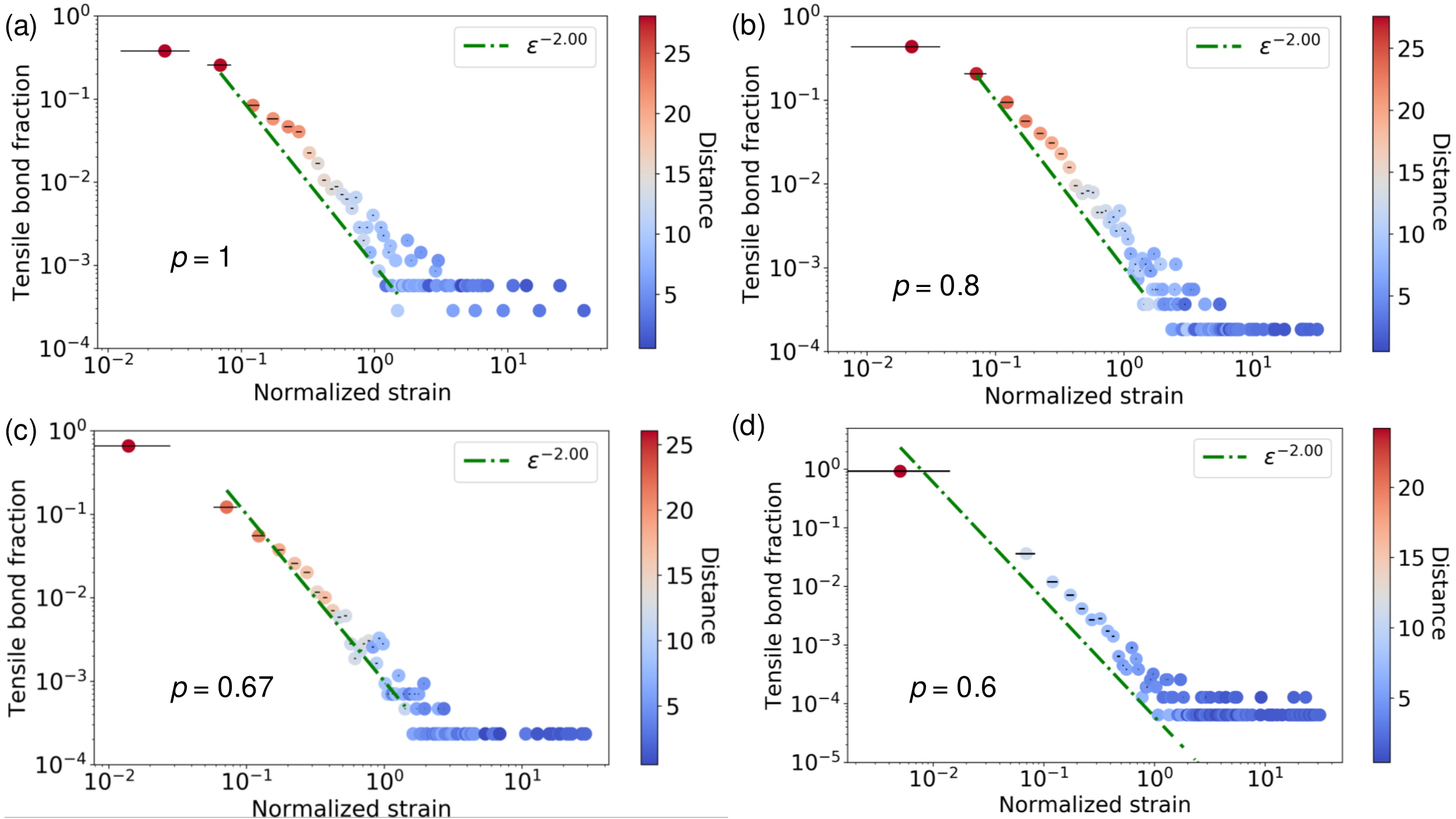}
            \caption{ \textbf{Strain distribution suggests bending modes decrease force transmission.} Each plot shows the normalized number of tensile bonds plotted against the corresponding bond strain (normalized by the threshold value), with colorbar indicating the mean distance of bonds in a given bin from the force dipole.
            (a) Strain distribution for a uniform network (all bonds present). (b,c) Strain histograms for Network 1 at $p = 0.8$ and $0.67$ respectively (Network 2, 3 and 4 also show similar results).
            Networks with $p = 1$ and $0.8$ are well within the stretching regime. Here the strain distributions show a continuous increase in strain as distance from dipoles decreases. However, for $p = 0.67$, the number of bonds in the lowest strain bin is higher than those for $p = 1$ and $0.8$. At the same time, the number of highly strained bonds is smaller. This suggests that bending reduces strains in this network (Network 2, 3 and 4 also show same result) and hints at the presence of a bend-stretch coupled regime.  
            (d) Strain distribution for four network realizations at p = 0.6. Bonds in these networks that are very close to the force dipole are highly strained while all other bonds have very little strain. The bonds at intermediate distances have very low strains because bending screens the long-range transmission of strain. These bonds lie in the first bin of the histogram with minimum strain and the mean distance of each bond in this bin is higher than 2.} 
            \label{fig:slope_strain_hist_bend}
        \end{figure*}



To further analyze the spatial decay of strains and to characterize the mechanical heterogeneity of these networks, we measure the strain distribution of tensile bonds. In Fig.~\ref{fig:slope_strain_hist_bend}, the fraction of stretched bonds is plotted against the corresponding value of tensile strain normalized by the threshold value, $\epsilon/\epsilon_{0}$. The color bar indicates the distance of the corresponding bonds from the center of the dipole, suggesting a continuous variation in bond strain with distance.
 We expect compressed bonds to be similarly distributed (not shown).
For networks in the stretching-dominated regime, $p>p_{cf} \approx 0.67$ shown in Figs.~\ref{fig:slope_strain_hist_bend}a-b, we expect the bond strains to closely follow that made by an affine deformation. As we go further in radial distance $r$ from the center of the dipole, the number density of available bonds at that radius increases as $n(r) \sim r$, while linear elasticity theory predicts that the strain decays as $\epsilon \sim r^{-2}$ ( Appendix B). Together, these predict a $n(\epsilon) \sim \epsilon^{-2}$  scaling of number of bonds with strain, shown as the dashed line in the strain distributions.  We see that the stretching-dominated networks follow this affine prediction very closely, except at very low strains ($\epsilon < 10^{-1}$)  corresponding to a large number of distant bonds without significant strain, and at high strains, corresponding to the few bonds very close to the dipole.  These account for only a few bonds per value of strain and thus do not contribute to the continuous $\epsilon^{-2}$ strain distribution. 

As we go through the stretching to bending-dominated transition (Fig.~\ref{fig:slope_strain_hist_bend}c-d),  the number of significantly strained bonds decreases.  In Fig.~\ref{fig:slope_strain_hist_bend}c, we show the strain distribution for Network 1 in the bend-stretch coupled regime ($p = 0.67$). Here, the number of  bonds with low strain (seen in the leftmost bin of the histogram) is comparatively more than the networks with $p = 0.8$ and $p = 1$. Other networks in this regime have similar histogram profiles as well. For bending-dominated networks at $p=0.6$ (Fig.~\ref{fig:slope_strain_hist_bend}d) and $p=0.48$ (Fig. S5), almost all bonds ($~ 92\% $) have very small strains in the $\epsilon/\epsilon_{0} < 10^{-2}$ bin. These represent all bonds beyond a distance of $\approx 10$ from the dipole, as seen from the color bar. We thus note a qualitative difference in the strain distributions of the bending-dominated networks. which show an absence of intermediate strains that are not very high or very low. This shows up as a noticeable gap in the range of normalized strain values between $\epsilon \sim 10^{-2} - 10^{-1}$. We show in SI Fig. S6 that this gap in strain values increases with increasing bond dilution.

Overall, we find that the stain energy decays as a  power law function of radial for the stretching-dominated ($p> 0.67$) networks, while  the bending-dominated ($p<0.67$) networks  do not follow a single power law. Increasing the depletion of bonds in networks leads to an increase in rate of decay of strain energy (Fig.~\ref{fig:Single dipole strain energy}). This is because the removal of bonds leads to bonds near the dipoles being strained more than in uniform networks (compare Figs.~\ref{fig:slope_strain_hist_bend} a and b). The strain distribution was quite homogeneous in stretching- dominated networks. However, we find an absence of intermediate strains in bending dominated cases. To further quantify the networks' response to applied contractile forces, and interaction between two force dipoles, we next study clusters formed by two dipoles.



\subsection{Strain clusters for two dipoles}

        \begin{figure*}[hbt]
            \centering
            \includegraphics[width=16cm]{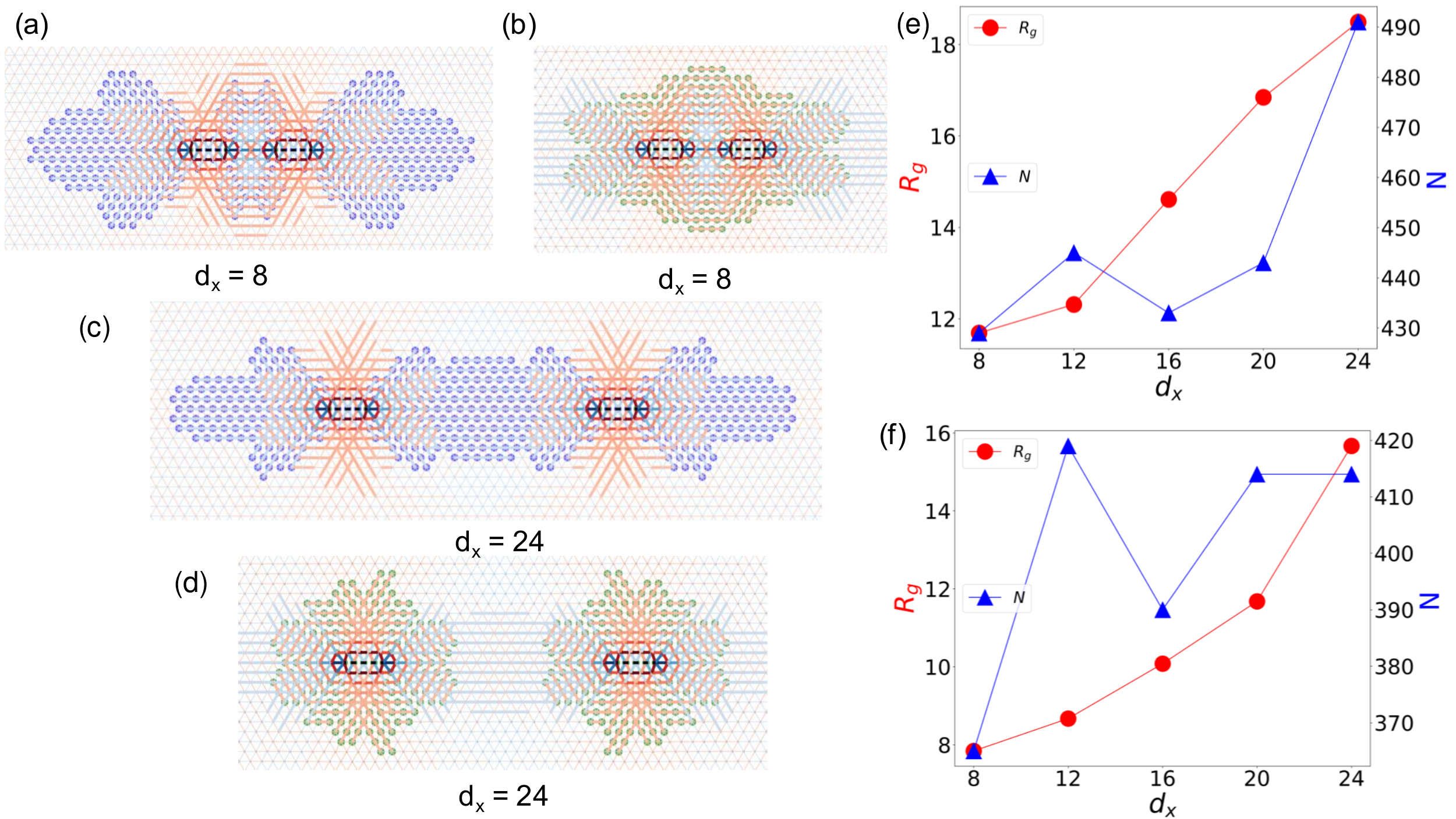}
            \caption{\textbf{Tensile and compressive cluster analysis for two dipoles in a uniform network} 
            (a) Tensile clusters formed by two dipoles with separation $d_{x} = 8$. (b) Compressive clusters formed by two dipoles with separation  $d_{x} = 8$. (c) Tensile clusters formed by two dipoles with separation  $d_{x} = 24$. Tensile clusters formed by two force dipoles are connected at all distances we can sample in our simulation box. (d) Compressive clusters formed by dipoles with separation  $d_{x} = 24$. When there is a small separation between the force dipoles, the compressive clusters form one large cluster, but separate on increasing distance between dipoles. (e) The size of the tensile cluster increases with increasing separation between the two dipoles. However, there is a small dip in the number of nodes as the separation increases from 12 to 16. As separation increases, more nodes that lie between the dipoles become available for cluster formation. At the same time, the extent of the cluster along the 
            $y-$axis decreases since there is less overlap between the tensile regions formed by the two dipoles. (f) The size of compressive clusters increases with increasing separation between the two dipoles. But, when the separation between dipoles is 16 and larger, the clusters are disconnected. This leads to a dip in cluster size at $d_{x} = 16$, after which at larger distances, the cluster size does not change as a function of separation.
            }
            \label{fig:Two dipole cluster}
        \end{figure*}

We now consider the combined network deformations by a pair of dipoles, which could be positioned in a variety of configurations.  This will help identify how the deformations by one dipole affect the other, and potentially elucidate long-range mechanical interactions between myosin motors in the cytoskeletal network.  We will consider two dipoles oriented along the $x-$axis, but which could be separated along their axes by a distance $d_{x}$, or transverse to their axes, by  a distance $d_{y}$. A third possibility, when the second dipole is rotated to be perpendicular with respect to the first, is shown in the SI Fig. S7.

In uniform networks with two dipoles placed along the $x-$axis, we find that tensile as well as compressive cluster size generally tends to increase with separation between dipoles (Fig.~\ref{fig:Two dipole cluster}). At close distances, there is a large region of overlap between the tensile strain clusters produced by both dipoles. As the distance increases, this region of overlap decreases in size leading to a higher net number of nodes that participate in the two-dipole cluster. There is also a secondary effect that changes the number of nodes in the tensile cluster with separation. At small $d_{x}$, the combination of two dipoles causes more bonds along the vertical direction to also become highly tensile. As $d_{x}$ increases, this effect decreases, and the cluster becomes more localized along the $x-$axis. The decrease in tensile bonds on the vertical axis and increase in tensile bonds between the two dipoles as $d_{x}$ increases, compete to decrease and increase the cluster size, respectively. It is this competition that presumably leads to a decrease of $N$ when the separation changes from $d_{x} = 12$ to $d_{x} = 16$ in tensile clusters seen in Fig.~\ref{fig:Two dipole cluster}e. At the farthest distance we sample, the size of the cluster is larger than twice the size of a tensile cluster formed by a single force dipole (Fig.~\ref{fig:Single dipole}). 

In the compressive case, the two dipoles form a unified single cluster at small separation (Fig.~\ref{fig:Two dipole cluster} b). However, as distance increases, the two clusters become disconnected (Fig. ~\ref{fig:Two dipole cluster interaction dx} b). This disconnect, whose onset is marked by a dip in the cluster size at $d_{x}=16$ in Fig.~\ref{fig:Two dipole cluster}f is followed by creation of two clusters that are independent of each other. The total number of nodes that make up these two clusters is approximately equal to twice the size of a compressive cluster formed by a single dipole. Due to this, $N$ reaches a maximum and does not change in value at $d_{x} = 20$ and $24$. This suggests that there is no significant interaction between the two dipoles at this distance. Combined with the size increase in tensile clusters, this shows that the two force dipoles interact through tensile bonds at large distances and compressed bonds do not play a role. 


\subsection{Two dipole cluster interactions}
        \begin{figure*}[htp]
            \centering
            \includegraphics[width=16cm]{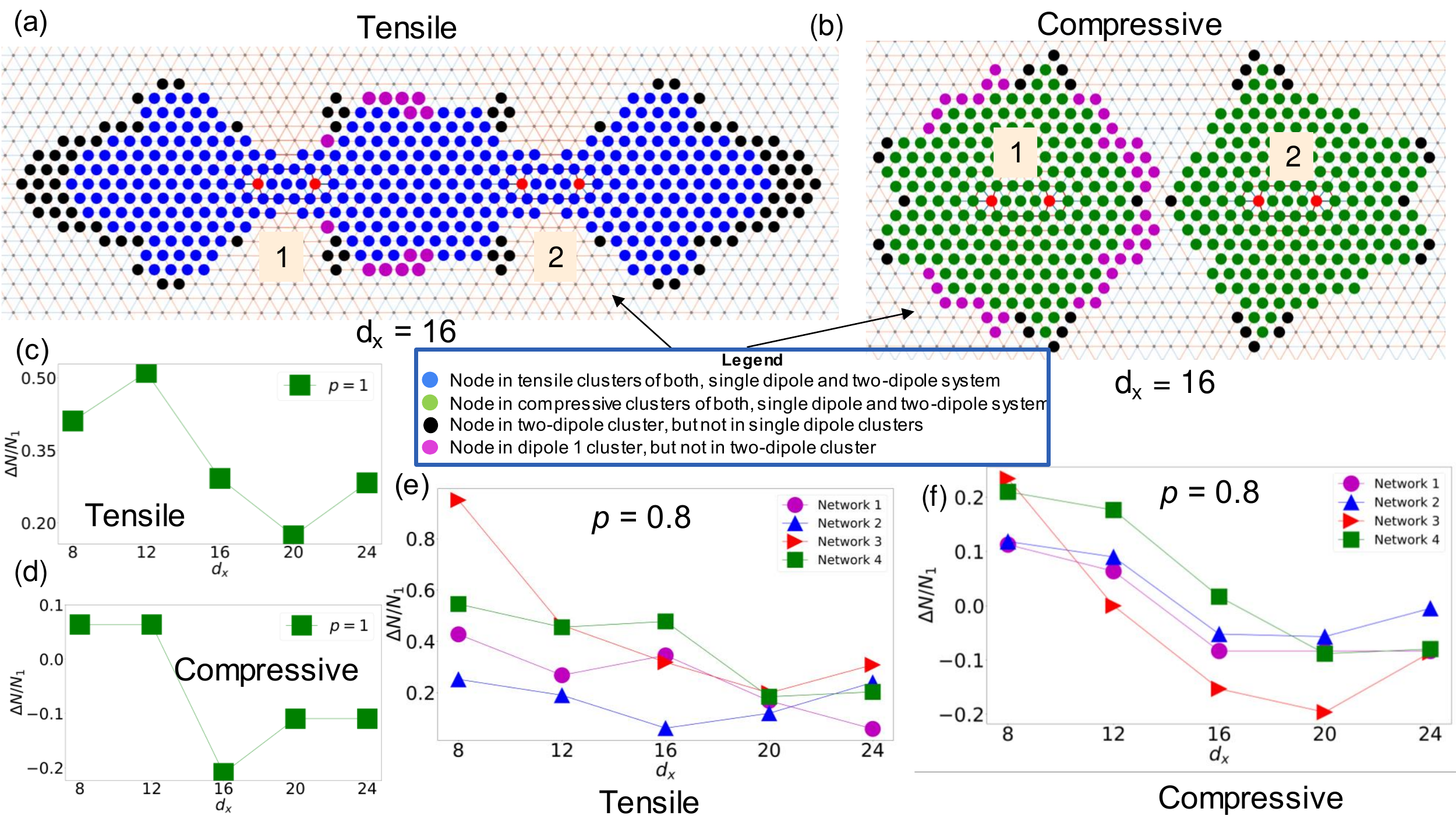}       
            \caption{ 
             {\textbf Cluster interaction analysis for two dipoles separated along $x-$axis in stretching-dominated networks. } 
            (a) Tensile cluster at dipole separation $d_{x} = 16$. Nodes that are part of the cluster formed by both dipoles but were not a part of the cluster made when only a single dipole was present are colored in black. Nodes that are not a part of the two dipole cluster but were a part of the cluster formed by dipole 1 (left) alone are colored in magenta. These 
            black and magenta colored nodes show how the cluster formed by the left dipole is influenced by the presence of the right dipole. The blue nodes occur in both the cluster formed by two dipoles and in one of the two single- dipole clusters. The presence of the right dipole 2 increases the extent of the tensile cluster of left dipole to its left along the $x-$axis. 
            (b) Compressive cluster formed by two dipoles at a separation of $d_{x} = 16$. The magenta nodes around dipole 1 on the left, show that the presence of dipole 2 on the right, reduces the  cluster formed by dipole 1  - suggesting an antagonistic effect that ``shields'' some nodes from the presence of dipole 1. 
            (c-f) Quantification of the effect of one dipole on the strain cluster of the other, $\Delta N$, as a function of separation between two dipoles. Here, $\Delta N = N_{12} - (N_{1} \cup N_{2})$ where $N_{12}$ is the number of nodes in the two-dipole cluster, while  $N_{1,2}$ are number of nodes in the cluster when only dipole 1 or dipole 2 is present.
            (c) In a uniform network ($p=1$), as the dipole separation increases, new nodes between the dipoles are available to form a combined tensile cluster, while there is a decrease in the combined stretching of the two dipoles, resulting in non-monotonic behavior.
            (d) $\Delta N$ for compressive clusters as a function of distance in a uniform network. A single shared cluster exists between the dipoles at $d_{x} = 8$ and $12$, while at $d_{x} \ge 16$, the cluster separates into two different clusters. The shielding shown in (b) leads to a reduced two-dipole cluster when compared to isolated single dipole clusters.
            (e) $\Delta N$ for tensile clusters as a function of distance in a disordered network in the stretching regime ($p=0.8$). In some networks, the cluster size made by two dipoles can be much larger than the sum of individual clusters of each dipole, indicating network-specific effects upon bond dilution. 
            (f) $\Delta N$ for compressive clusters as a function of distance in disordered networks in the stretching regime ($p=0.8$) shows a similar behavior seen for compressive clusters in a uniform network in (d), with the dip corresponding to cluster separation. 
            }

            \label{fig:Two dipole cluster interaction dx}
        \end{figure*}
We now aim to quantify how the presence of a second dipole modifies the tensile and compressive force clusters created by the first dipole. To do so, we consider simulations performed for three cases: dipole 1 (left) alone, diple 2 (right) alone, and both dipoles 1 and 2 present  (Fig. ~\ref{fig:Two dipole cluster interaction dx}). We focus on how the presence of the dipole 2 on the right modifies the cluster around dipole 1 on the left.  Equivalently, we could have considered the effect of dipole 1 on the cluster around dipole 1, but in general this could be different because of the difference in local network structure around the dipoles in the bond-diluted cases. However, since we present these results for slightly diluted ($p>0.8$) networks, these fluctuations are expected to be small. This expectation is supported by our observation of similar trends in four different specific realizations of the diluted networks.  The fluctuations are expected to grow for bending-dominated networks, which we do not consider in the current analysis, because the cluster of strained bonds is very small in these cases (Fig.~\ref{fig:Single dipole bend}).

        \begin{figure*}[htp]
            \centering
            \includegraphics[width=16cm]{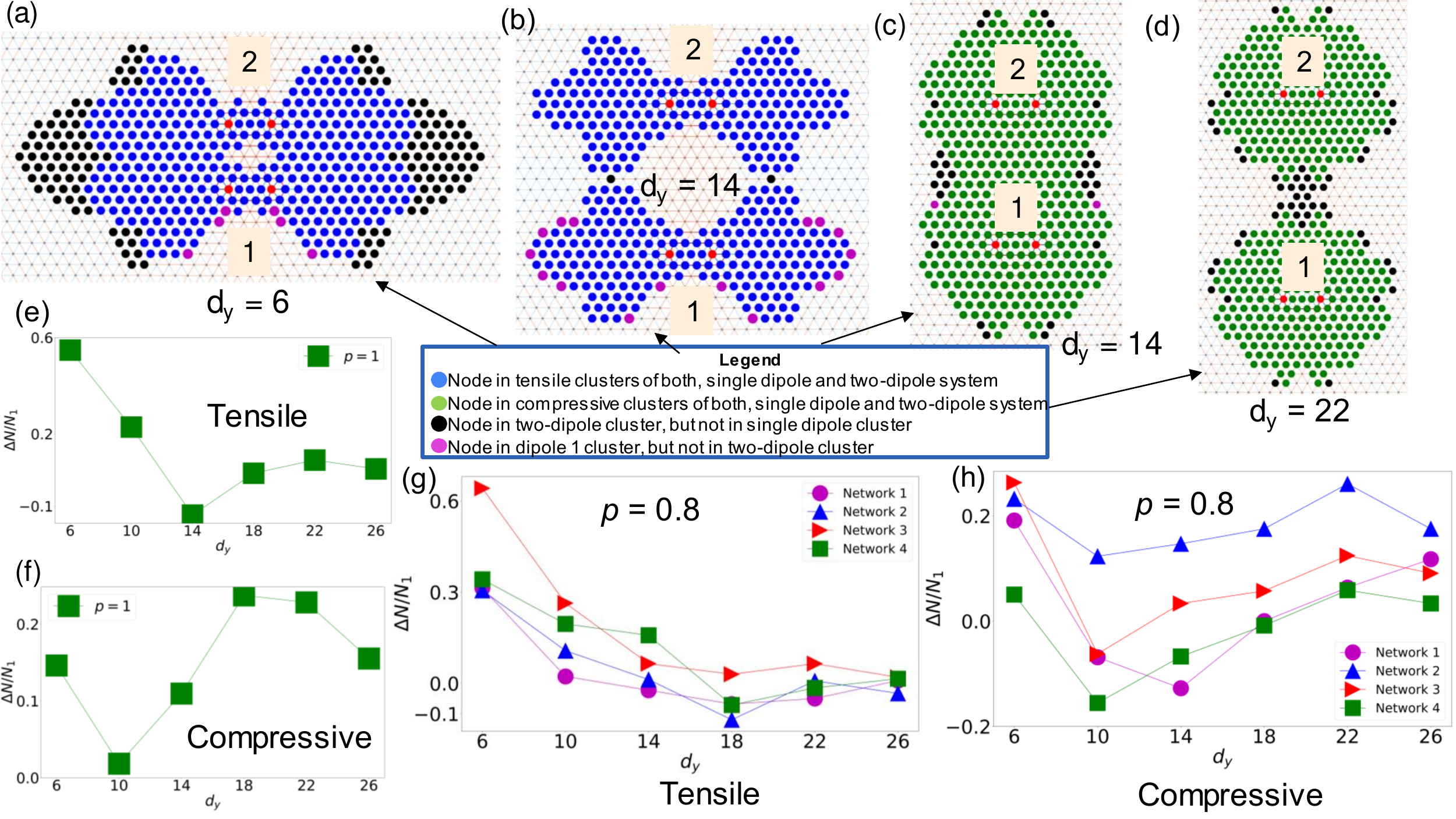}           
            \caption{ \textbf{Cluster interaction analysis for two dipoles separated along $y-$axis in stretching-dominated networks.} 
            (a) When the dipoles are separated by 8 rows on the $y-$axis, the two dipoles make one connected large tensile cluster. The black nodes show that this cluster has many nodes that were not a part of the tensile cluster formed by the individual dipoles, thereby indicating a reinforcing interaction between two dipoles. 
            (b) The two dipoles have a single connected cluster even when y separation increases to 14 rows. For larger distances, the cluster divides into two disconnected clusters.
            (c) Two dipoles when separated by 14 rows form one large cluster.
            (d) At a separation of 22 rows, the two dipoles still form one connected cluster. However, the region of overlap is small. Beyond this distance, the two dipoles form two separated clusters.
            (e-h) Measure of cluster interaction vs separation distance between two dipoles: $\Delta N = N_{12} - (N_{1} \cup N_{2})$ where $N_{12}$ is the cluster size when two dipoles are present and $N_{1,2}$ are cluster sizes when only one of either of the dipoles is present.
            (e) 
            For tensile clusters in uniform network, $\Delta N$ decreases with increasing separation 
            . This is because tensile regions are found primarily to the left and right of the dipoles and the region of overlap of tensile cluster decreases with increasing $y$ distance between the two dipoles.
            After the cluster separates into two clusters for $d_{y} \ge 18$, $\Delta N$ does not show much variation.
            (f)  For compressive clusters in  uniform network, $\Delta N$ shows a 
            non-monotonic behavior. As the separation increases, new nodes between the two dipoles become available for the cluster. However,  there is also a decrease in combined compression of the two dipoles which leads to this non-monotonic behavior.
            (g,h) Tensile and compressive clusters for two dipoles separated along $y-$axis in disordered networks in stretching regime show similar trends as seen in the corresponding plots of a uniform network (e,f).
            } 
            \label{fig:Two dipole cluster interaction dy}
        \end{figure*}

In Fig. ~\ref{fig:Two dipole cluster interaction dx}, nodes are colored differently depending on whether they belong to the cluster arising in a single dipole or a two-dipole simulation. For concreteness, let us define $C_{12}$ as the set of nodes in the tensile cluster when both dipoles 1 and 2 are present, while $C_{1}$ and $C_{2}$ are  clusters when only dipole 1 or only dipole 2 is present, respectively. 
 The blue nodes shown in (Fig. ~\ref{fig:Two dipole cluster interaction dx} a) are common to the cluster formed by dipole 1 or dipole 2 alone, as well as to the cluster formed by the combined effect of both dipoles 1 and 2: $(C_{1}\cup C_{2}) \cap C_{12} $.  The black nodes belong to the combined cluster of both dipoles ($C_{12}$), but are not present in single dipole clusters induced by dipole 1 or dipole 2:  $C_{12} -(C_{1} \cup C_{2})$. So the black nodes show that cluster size increases due to interaction between the two dipoles. Their number is a measure of the extent of positive interaction or reinforcement between the two dipoles . Magenta nodes belong to the cluster induced by dipole 1, but are not present in the combined cluster of the two dipoles: $C_{1} - C_{12}$. Therefore, the number of magenta nodes is a measure of the negative interaction or shielding effect of dipole 2 (right) on the cluster of dipole 1 (left). 
 
 Tensile clusters induced by two dipoles separated by $d_{x} = 8$ (Fig. ~\ref{fig:Two dipole cluster interaction dx}a) show positive reinforcement in cluster size due to the presence of the the right dipole (more black nodes than magenta). However, compressive clusters made by the two dipoles at the same separation  show almost no positive reinforcement (Fig. ~\ref{fig:Two dipole cluster interaction dx}b). Instead, the magenta nodes are more abundant than the black nodes. Thus, the right dipole seems to shield nodes that would have been a part of the compressive cluster of the left dipole. To quantify the interaction, we calculate the difference in the number of nodes that belong to the single cluster of the two-dipole system and the number of nodes that occur in the cluster of dipole 1 alone or dipole 2 alone: $\Delta N = n(C_{12}) - n(C_{1} \cup C_{2})$. This quantity, normalized by the corresponding number of nodes in a single dipole cluster, is a measure of the positive or negative interaction between two dipoles.  Tensile clusters show positive interference ($\Delta N >0$) for all separations along the $x-$axis (Fig. ~\ref{fig:Two dipole cluster interaction dx} c). However, $\Delta N$ for compressive clusters shows both positive and negative interference depending on separation distance (Fig. ~\ref{fig:Two dipole cluster interaction dx} d). At a separation of $d_{x} = 16$ along the $x-$axis, the dipoles make two separate compressive clusters (Fig. ~\ref{fig:Two dipole cluster interaction dx} b), instead of a single large cluster formed when they are closer such as at $d_{x} =8$. see Fig. 8b. The shielding of the dipole 2 (right) on the cluster formed by the dipole 1 (left) is clearly seen as the magenta nodes in Fig. ~\ref{fig:Two dipole cluster interaction dx} b. This shielding effect is maximum when $d_{x}=16$ as shown by the minimum in $\Delta N/N_{1}$ in Fig. ~\ref{fig:Two dipole cluster interaction dx}d and decreases as the two dipoles move further apart. Interaction between dipole strain clusters in slightly depleted networks that are in stretching regime ($p=0.8$) show similar trends in $\Delta N$ as uniform networks (Fig. ~\ref{fig:Two dipole cluster interaction dx} e,f).

We also similarly quantify the effect of dipole 2 (on the top) on the cluster formed by dipole 1 (on the bottom), when the two dipoles are separated along $y-$axis (Fig. ~\ref{fig:Two dipole cluster interaction dy}). Tensile clusters show positive interference for short distances and $\Delta N$, represented by black nodes, is positive for $d_{x} = 6, 10$ (Fig. ~\ref{fig:Two dipole cluster interaction dy} e). However, at $d_{x} > 14$, the two dipoles form two separate tensile clusters instead of a combined cluster, and the $\Delta N$ drops close to $0$. This is different from the case of separation of the two dipoles along their axis, where the tensile clusters remain connected, even at large distances of separation, and $\Delta N$ is substantially greater than $0$ (Fig. ~\ref{fig:Two dipole cluster interaction dx} a,c,e). As opposed to compressive clusters formed by dipoles separated along $x-$axis (Fig. ~\ref{fig:Two dipole cluster interaction dx} d), $\Delta N$ for compressive clusters when the dipoles are separated along $y-$axis shows only positive interference (Fig. ~\ref{fig:Two dipole cluster interaction dy} f). $\Delta N$ for slightly diluted networks shows a behavior similar to that found in uniform networks (Fig. ~\ref{fig:Two dipole cluster interaction dy} e). However, $\Delta N$ for compressive clusters in diluted networks for dipoles separated along $y-$axis shows negative interference (Fig. ~\ref{fig:Two dipole cluster interaction dy} h), which was not seen in the case for uniform networks. Moreover, compressive clusters show positive reinforcement when dipoles are separated along the $y-$axis (Fig. ~\ref{fig:Two dipole cluster interaction dy} c,d,f,h) while clusters formed by dipoles separated along $x-$axis showed a more pronounced shielding effect (Fig. ~\ref{fig:Two dipole cluster interaction dx} b,d,f).


        \begin{figure*}[htp]
            \centering
            \includegraphics[width=16cm]{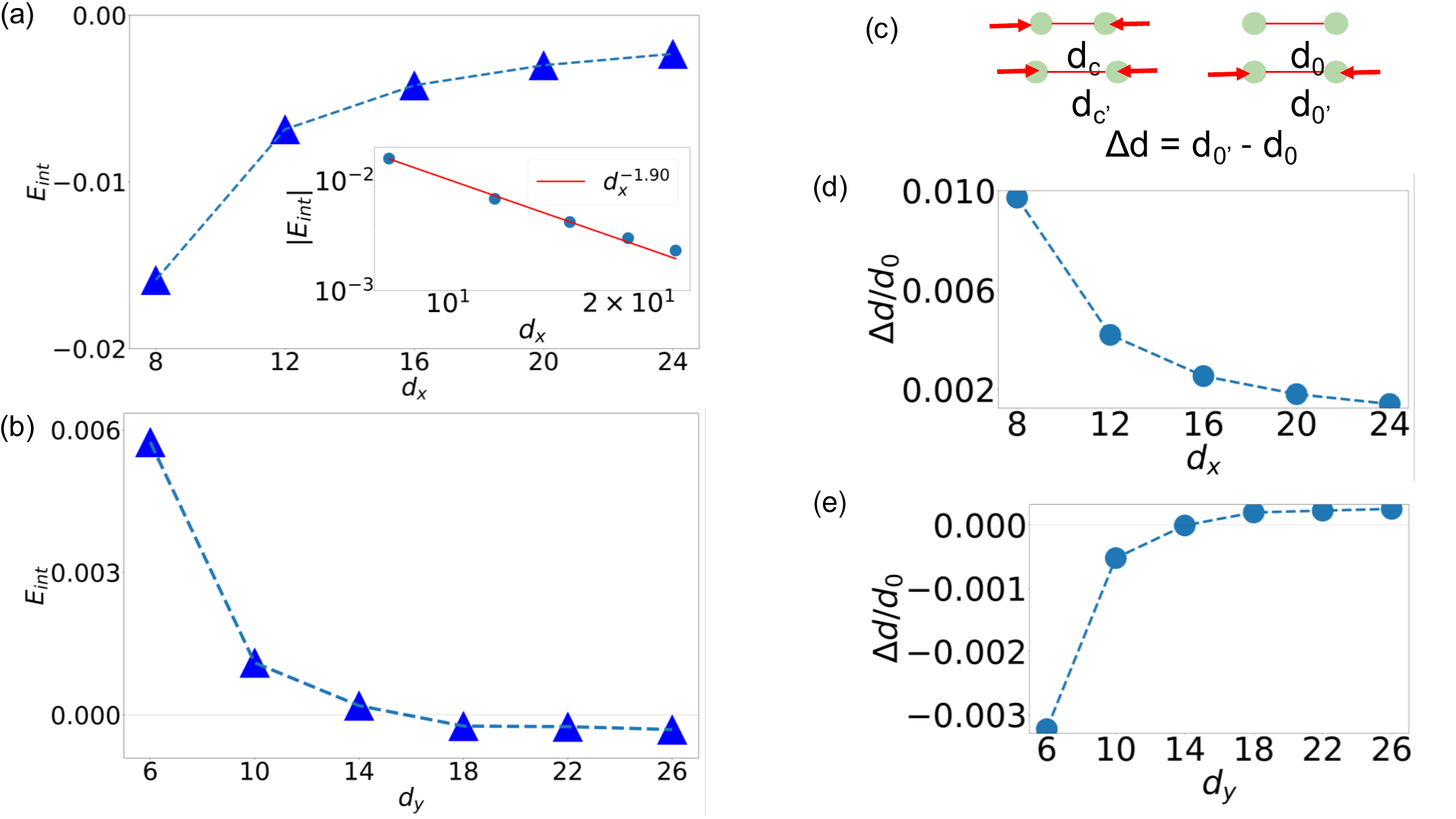}
            \caption{\textbf{Interaction between a pair of dipoles in a regular ($p = 1$) lattice.} 
            The interaction energy (a,b) is the extra elastic energy of deformation in the fiber network when both dipoles are present in comparison to when only one of them is present. This is a global measure of interaction between two dipoles. This interaction is negative when the two dipoles are separated along the $x-$axis(a). This indicates an attractive force between these dipoles. The inset shows the fit of a power law which is close to the relation $|E_{int}| \sim d^{-2}$, expected for interaction between dipoles in 2D from linear elasticity theory. (b) The interaction energy between dipoles when separated along the $y-$axis is positive, indicating a repulsive interaction between them. 
            (c) A schematic diagram that shows a local measure of interaction between the two dipoles, corresponding to how much a dipole node displaces in the presence of the other dipole. (d,e) The measured values of local dipole interaction for separation along $x$-axis and $y$-axis respectively. As expected, when the extra dipole is placed to the right of the central dipole, in a region with predominantly extensile bonds, $\Delta d$ is positive. The opposite is true when the extra dipole is placed above the central dipole on the $y-$axis, where the central dipole produces a region of predominantly contractile bonds. In both cases, the strength of the interaction decreases with increasing distance.
            }
            \label{fig:Two dipole interaction energy}
        \end{figure*}
        
\subsection{Two dipole interaction energy}   

        \begin{figure*}[htp]
            \centering
            \includegraphics[width=16cm]{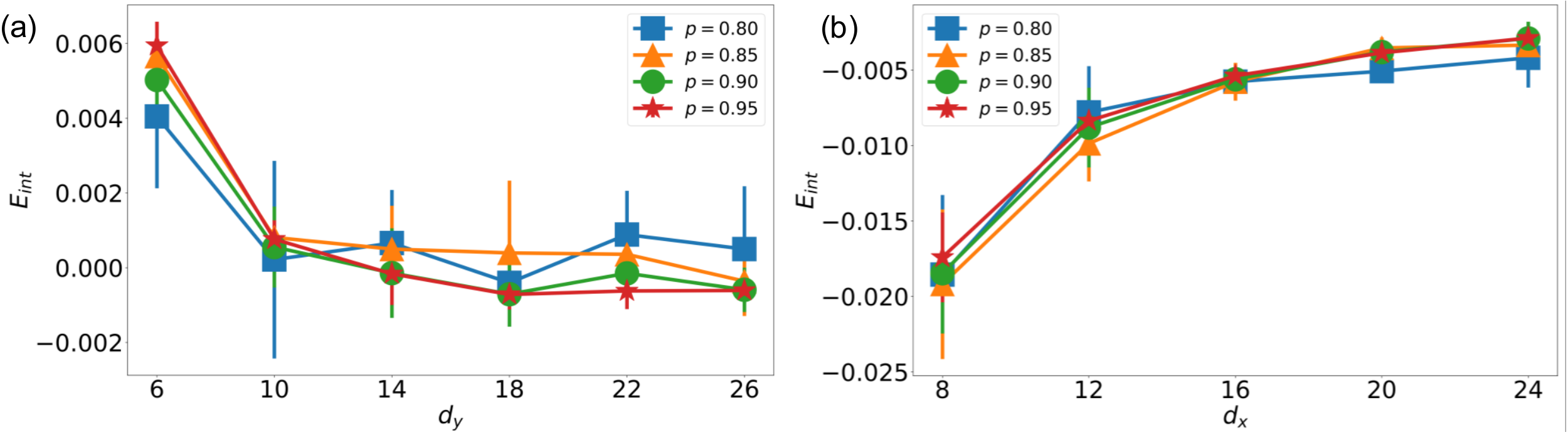}
            \caption{ \textbf{Interaction energies in depleted networks in stretching regime.} (a) The interaction energies for dipoles separated along the y axis suggests a repulsive force between the two dipoles which gets weaker as the separation increases. (b) The interaction energy between two dipoles separated along x axis suggests an attractive force between the dipoles which decreases as the separation increases.} 
            \label{fig:interaction energy stretch}
        \end{figure*}

In analogy with electric charge or defects in an elastic medium, the interaction energy for a given configuration of two force dipoles is the extra elastic energy of the medium when both dipoles are present, in comparison to when only one of them is present. We calculate this interaction energy using,
\begin{equation}\label{eq:int energy}
E_{\mathrm{int}}(0,{\bf d}) = E_{12} (0, {\bf d}) - E_{1} (0) - E_{2} ({\bf d})
\end{equation}
where $E_{12}$ is the total elastic energy of the network with both dipoles at a prescribed separation ${\bf d}$.  $E_{1} (0)$ and $E_{2} ({\bf d})$ represent the total elastic energy of the network when dipole 1 alone is present at the origin, and when dipole 2 alone is present at a position ${\bf d}$, respectively. Since we consider a triangular lattice with periodic boundary conditions, the system is translationally invariant and only the relative separation of the two dipoles matters. A negative (positive) value of $E_{\mathrm{int}}$ indicates a favorable (unfavorable) interaction between the two dipoles.

In  Fig.~\ref{fig:Two dipole interaction energy}a-b, we placed two dipoles exerting contractile stress along the $x$-axis in a uniform network, and varied their relative separation - first, along the $x-$axis and then, along the $y-$axis.  We find that when the dipoles  are separated along the $x-$axis, the interaction is favourable with negative interaction energy values. The interaction energy also weakens as the separation between the dipoles increases, and is expected to tend to zero for infinite dipole separation. The negative interaction energy occurs as a result of the second contractile dipole being placed in the region of the network that is stretched by the first dipole (Fig. 2a). The elastic energy of some of the stretched bonds is therefore lowered in the region of the second dipole.  We have seen before (Fig.~\ref{fig:Two dipole cluster}) that tensile clusters mediate much of the interaction between two dipoles at large distances in a uniform network. However, at small distances, both tensile as well as compressed bonds mediate the interaction.  When we separate the dipoles along the $y-$ axis, the interaction is not favorable and the interaction energies are positive - indicating a repulsive force between the dipoles. In these networks, compressive clusters mediate much of the interaction between the two dipoles even at large distances, while tensile clusters are clearly separated for cases of dipoles at large distances (Fig. ~\ref{fig:Two dipole cluster interaction dy} b,c).  Of note, the decay of the interaction energy for dipoles separated along their axis follows the $d^{-2}$ behavior predicted by continuum elasticity theory, whereas dipoles placed transverse to their axis lack such a regular trend. This behavior of interaction energy is not dependent on the direction of the dipole forces being along a lattice symmetry direction, and is also seen for dipoles aligned along the $y-$ axis (not a lattice symmetry direction) as confirmed in Fig. S8.

To obtain a local measure of interaction between the dipoles, we also calculate the difference in spacing of the nodes of the central dipole caused by the extra dipole (Fig.~\ref{fig:Two dipole interaction energy}). The extra dipole, when placed in a stretched region (to the right of the central dipole,) shows that this value is positive and decreases as the separation between the two dipoles increases. The behavior is opposite when we place the extra dipole on the $y-$axis in a contractile region right above the central dipole. We also calculate the interaction energy in slightly-depleted networks that lie in the stretching-dominated regime (Fig.~\ref{fig:interaction energy stretch}. These show a trend similar to that seen in uniform networks in Fig.~\ref{fig:Two dipole interaction energy}. In the bending-dominated regime, the fluctuations in interaction energy are too strong to give any trends. We thus show, that unless strongly depleted, pairwise dipole interactions exhibit regular trends with mutual orientation and separation, which predicts favorable alignment of force dipoles along their axis.






\section{Discussion}
Elastic fiber networks are ubiquitous in synthetic and biological materials. Biopolymer networks such as actin in the cellular cytoskeleton or collagen and fibrin in the extracellular matrix of tissue are subject to mechanical stresses - both external loading and internal forces actively generated by molecular motors. In response to such forces, these fibrous materials exhibit unique, non-linear mechanical properties that are crucial to their biological function and competing demands - such as the ability to remodel as well as to preserve integrity \cite{Burla2019}. Even if lacking the full molecular complexity and structural hierarchy of biomaterials, elastic fiber network models such as the one considered here, capture essential aspects of their mechanical properties, such as an abrupt stiffening transition under shear \cite{sharma2016strain} and long-range force transmission \cite{Ronceray16, Sopher2018, alisafaei2021}.  In the present work, we not only addressed how strain propagates through such a model elastic network from a force dipole representing, for example, molecular motor activity, but also investigated how two such dipoles may interact through the strains that they generate. 

In the first part of this work, we explored the range and heterogeneity of force transmission from a single local force dipole, in the elastic network.  We deployed several metrics to quantify the extent of force transmission: the size (number of nodes, radius of gyration) and shape of connected clusters of stretched and compressed bonds, the decay of strain energy, and the distribution of strains in different bonds.  We showed that these metrics depend on two key elastic network parameters: the bond dilution probability and the dimensionless bending-to-stretching stiffness ratio. Prior studies have shown how the macroscopic response of such bond-diluted elastic networks to external shear depends on these parameters. In particular,  under-coordinated networks ($p < p_{CF}$) with $\widetilde{\kappa} \ll 1$ show a bending-dominated response characterized by floppy modes consisting of easily rotating bond clusters. 
In this work, we examine how the force clusters around single dipoles are modified under bond dilution.

While buckling under compression is a generic feature of slender fibers, we here considered fiber bending in response to transverse forces alone.  For real networks, this may correspond to having smaller dipole forces or thicker, laterally cross-linked bundles of fibers.  In networks with fiber buckling, bonds transverse to the dipole axis will be under compression leading to buckling and softening, such that the tensile force is focused along longer, force chains. The asymmetry of fibers under tension and compression due to buckling also results in different force distributions behaving as effectively contractile at larger scales \cite{Ronceray16}.  We show that bending alone gives the opposite trend, decreasing the range of force transmission. When floppy bending modes are available, most fibers will respond to dipole force by bending through bond rotations instead of bond stretching. This screening of bond strains by bending results in an anomalous shortening of the range of force transmission in the network, as shown by the decay of elastic energy with distance from the dipole in Fig. ~\ref{fig:Single dipole strain energy}. While over-coordinated networks result in clear power law decays of elastic strains with distance, as predicted for affine deformations, the strain decay in bending-dominated networks did not follow any clear power law. Overall, this suggests that bending and buckling have opposite effects on the range of force transmission.  How these opposite trends compete in networks that allow easy bending is an interesting question for future study.

While previous works have explored mechanical interactions between two isotropic force distributions representing cells in an extracellular matrix \cite{mann2019force}, this is the first exploration of analogous effects for a pair of anisotropic force dipoles representing the contractility of myosin motors in the actin cytoskeleton, or at a different scale, between two polarized cells in a fibrous extracellular medium.  Recent evidence from cell biology suggests that such long-range mechanical interactions between myosin motors may drive them into spatial registry across stress fibers \cite{hu_17, dasbiswas_19}.  In general, mechanical interactions through an elastic medium may direct the self-organization of the cell and tissue into ordered, functional structures such as registered fibrils in muscle tissue \cite{dasbiswas_15} or multicellular networks of endothelial cells \cite{noerr2022}. While many experiments that demonstrate mechanical interactions between cells are carried out on linear elastic hydrogel substrates, natural biomaterials in the extracellular matrix or cytoskeleton typically occur as fiber networks that are strongly nonlinear in their mechanical response. Further, such disordered networks transmit forces heterogeneously at the scale of individual fibers that cannot be captured by continuum elastic models. Over long time scales the cytoskeletal network also undergoes significant remodeling and can exhibit fluid flow \cite{moeendarbary_13} - such viscoelastic or poroelastic effects are not considered in the present study which focuses on the short time scale elastic interactions.

Here, we quantified the elastic interactions that may arise between two distant force dipoles embedded in a fiber network using different metrics.  We showed that they differently affect the sizes and shapes of each other's strain clusters, which can be considered to be their ``regions of influence'', depending on their relative position and orientation. Clusters comprising stretched or compressed bonds also showed qualitative differences. Specifically, when separated along their principal axis, one dipole reinforced stretching due to the other, but reduced the overall compression. The elastic interaction energy between two dipoles followed trends predicted by linear elasticity theory for a uniform network, as expected (Fig. ~\ref{fig:Two dipole interaction energy}).  In particular, the two dipoles resulted in an energetically favorable (''attractive'') configuration when separated along their principal axes, but resulted in an unfavorable (''repulsive'') configuration when separated in the transverse direction. These trends were preserved in networks where a small amount of disorder was introduced ($p>0.8$). Both these results for elastic interaction energy and reinforced tensile bonds in two-dipole systems suggest that mechanical interactions between similarly oriented actomyosin units may lead to their lining up to form a stress fiber. Recent experiments do show that stress fibers are built up from the initially disordered cytoskeleton through the contractile myosin motor activity \cite{Lehtimaki2021}.  However for diluted networks ($p < 0.6$), it was not possible to obtain such general trends. The elastic deformation energy is very sensitive to the local network structure around the dipole, and differs strongly from one network configuration to another. Such strong strain and elastic energy fluctuations lead to the loss of any trends on the average.  This may suggest that once locally dense, over-coordinated (corresponding to greater bond probability)  or strongly bundled regions (corresponding to greater bending stiffness) arise in the network, such as through cross-linking by actin binding proteins, mechanical interactions may drive the actomyosin units towards alignment into ordered structures. Such locally denser or ``patchy'' fiber network configurations have been recently shown to modify the rigidity percolation threshold  \cite{MichelPhysRevResearch2022}. Correlated fiber patches are likely to confer additional stability to diluted networks leading to longer-range force transmission.  In conclusion, our work shows that elastic interactions can arise between distant force dipole in disordered, fibrous media, and that their strength and range can be enhanced by suppressing fiber bending.

    




\section*{Appendix}
\subsection*{Appendix A: Fiber network model energy and forces}

We calculate for the whole network, stretching, bending and dipole energies, given in  Eq.~\ref{eq:total energy_main}.  The network configuration at mechanical equilibrium is obtained numerically by minimizing the total energy.

The stretching energy is given by a pairwise sum over nodes,
\begin{equation}\label{eq:stretch energy}
E_{s} = \frac{\mu}{2} \sum_{\langle ij \rangle} (r_{ij}-r_{0})^2,   \tag{A1}
\end{equation}
where  $i$ and $j$ represent adjacent nodes that are connected by a bond of rest length, $r_{0}$ (set everywhere to $1$ in our simulations).  $r_{ij}$ is the actual bond length after force dipoles have been applied in the networks, 

Fiber bending is represented by the relative change in angle between two collinear bonds. These bonds connected three nodes denoted by $j$, $i$ and $k$, with $i^{th}$ node being central.  The total bending energy is given by the sum over all such node triplets wherever connected by bonds,
\begin{equation}
E_{b} = \frac{\kappa}{2 r_{0}}  \sum_{\langle jik \rangle}{2 \sin^2 (\theta_{jik}/2)},  \tag{A2}
\label{eq:bending_energy}
\end{equation}
with bond angle given by
\begin{equation}
\sin \theta_{jik} = \frac{|\boldsymbol{r}_{ij} \times \boldsymbol{r}_{ik}|}{|\boldsymbol{r}_{ij}| |\boldsymbol{r}_{ik}|}   \tag{A3}
\label{eq:sin_theta}
\end{equation}
Here $\boldsymbol{r}_{ij}$ and $\boldsymbol{r}_{ik}$ are the separation vectors connecting nodes $i$ to $j$, and nodes $i$ to $k$, respectively.

The dipole energy is the scalar product of force applied to and distance between nodes of the dipole.
\begin{equation}
    E_{d} = \sum_{\langle mn \rangle}\boldsymbol{F} \cdot {\boldsymbol{l}_{mn}}   \tag{A4}
\end{equation}
Here $\boldsymbol{l}_{mn}$ is the distance between $m^{th}$ and $n^{th}$ nodes that belong to the force dipole, not necessarily adjacent. Force is always along the separation vector between dipole nodes $m$ and $n$.  The total energy  $E_{t}$ is the sum of stretching, bending and dipole energies given above.




The stretching force on the $i^{th}$ node due to the $ij$ bond spring is given by the derivative of the stretching energy with respect to node position, and results in a central force, 
\begin{equation}
\boldsymbol{F}^{\langle ij \rangle}_{s,i} = -\mu (r_{ij} - r_{0}) \frac{\boldsymbol{r}_{ij}}{r_{ij}}   \tag{A5}
\end{equation}
The force due to collinear $ji$ and $ik$ bonds bending on the central $i^{th}$ node (in 2D) is given by the derivative of the relevant bending energy term in Eq.~\ref{eq:bending_energy} with respect to  displacement in the position of the $i^{th}$ node. 
\begin{equation}
    \boldsymbol{F}^{\langle jik \rangle}_{b,i} = \frac{2 \kappa}{r_{0}} \frac{\sin \theta_{jik}}{\cos \theta_{jik}} 
    \frac{\partial \sin \theta_{jik}}{\partial \boldsymbol{r}_{i}},   \tag{A6}
 \label{eq:bending_force_1}   
\end{equation}
where the gradient of the sine of the bending angle can be evaluated from Eq.~\ref{eq:sin_theta} as,
\begin{equation}
\begin{split}
    \frac{\partial \sin \theta_{jik}}{\partial \boldsymbol{r}_{i}} &=
     sgn(\hat{z} \cdot (\boldsymbol{r}_{ij} \times \boldsymbol{r}_{ik})) \frac{\hat{z} \times \boldsymbol{r}_{jk} }{|\boldsymbol{r}_{ij}| |\boldsymbol{r}_{ik}|} \\
&\quad
    - \sin \theta_{jik} \frac{\boldsymbol{r}_{ij}}{r_{ij}}
     - \sin \theta_{jik} \frac{\boldsymbol{r}_{ik}}{r_{ik}}      
\label{eq:gradient_sin}
\end{split} \tag{A7}
\end{equation}
where $sgn(x) = x/|x|$ represents the sign of the argument. The forces on the side nodes, $j$ and $k$, due to the bending of this angular spring at the central $i^{th}$ node, are similarly evaluated, as 
\begin{equation}
    \begin{split}
    \boldsymbol{F}^{\langle jik \rangle}_{b,j} &= \frac{2 \kappa}{r_{0}} \big( \tan \theta_{jik}  \frac{\partial \sin \theta_{jik}}{\partial \boldsymbol{r}_{j}}\big),\\
     \boldsymbol{F}^{\langle jik \rangle}_{b,k} &= \frac{2 \kappa}{r_{0}} \big( \tan \theta_{jik}  \frac{\partial \sin \theta_{jik}}{\partial \boldsymbol{r}_{k}}\big).  
     \end{split}  \tag{A8}
\end{equation}
Thus, every angular spring applies forces at three nodes.  The total bending force at the $i^{th}$ node will then involve sets  of three connected, collinear bonds passing through $i$, given all relevant bonds are present).
This calculated force is used to displace each node in the numeric conjugate gradient procedure to find the local energy minimum configuration.



\subsection*{Appendix B: Continuum elastic response to force dipole in 2D}

The uniform network, with all bonds present, undergoes affine deformations in response to imposed shear forces. This response is similar to that of a continuum elastic medium in 2D whose shear and bulk moduli are related to the stretching stiffness of each individual bond, $\mu$. We give here the expected deformations of an isotropic and homogeneous linear elastic medium in response to a single force dipole.  The stretching-dominated, affine, network behavior is expected to be closely approximated by this continuum model.

The displacement at a point ${\bf x}$ caused by a force acting at another point (chosen, without loss of generality, to be the origin) in a direction $j$  on the surface of an infinite linear and isotropic elastic medium in 2D is given by the appropriate Green's function \cite{landau_lifshitz_elasticity},
\begin{equation}
\begin{split}
u_{i}({\bf x}) = G_{ik}(\mathbf{x}) F_{k} 
&= \frac{1+\nu}{4 \pi Y}  \bigg((\nu-3) \delta_{ik} \log \bigg(\frac{|\mathbf{x}|}{a}\bigg) + \\
&\quad 
    (1+\nu) \frac{x_{i}x_{k}}{|\mathbf{x}|^2}  \bigg) F_{k} ,
\label{displacement_monopole}
\end{split}  \tag{B1}
\end{equation}
where $u_{i}({\bf x})$ is the displacement in the $i$th direction of the medium at point ${\bf x}$ caused by the $j$th component of the force $F$ at the origin, and the relevant elastic constants are the  2D stretching modulus $Y$ and Poisson's ratio, $\nu$, of the elastic medium.  

If instead of a point force, there is a pair of equal and opposite forces that are separated by a small distance (corresponding to the contractile actomyosin force dipole denoted by ${\bf P}$) the displacement is related to the derivative of the expression in the right hand side of Eq.~(\ref{displacement_monopole}) with respect to a spatial coordinate. The resulting, relative deformation of the elastic medium is given by the strain, which is a derivative of the displacement $u_{i}({\bf x})$ in Eq.~(\ref{displacement_monopole}), $u_{ij}({\bf x}) = P_{jk} \partial_{j} \partial_{k} G_{ik}$, where usual Einstein summation convention is implied. 

For a dipole aligned along the x-direction, only the $P_{xx}$ component is present. The decay of strain with distance due to a dipole can be easily seen as power counting. For an isotropic distribution of dipoles, the deformation depends on $G_{ii} \sim Y^{-1} \log (|x|/a)$. The direction-averaged trace of the strain goes as,
\begin{equation}
u_{ii} \sim \partial^{2}_{kk} G_{jj} \sim \frac{P}{Y} \frac{1}{r^{2}}  \tag{B2}
\label{strain_trace}
\end{equation}
and the corresponding elastic deformation energy density, $\varepsilon \sim Y u_{ii}^{2} \sim P^{2} Y^{-1} r^{-4}$.
This explains the observed trend in the decay of the direction-averaged strain energy as a function of distance observed in Fig. 6.

The interaction energy between two dipoles considered in Fig. 11 can be similarly derived within the framework of elasticity theory. It is the work done by a dipole, $\mathbf{P}^{\alpha}$ in deforming the substrate in the presence of the strain created by a second dipole $\mathbf{P}^{\beta}$, and  is given by \cite{Bischofs2004}, 
\begin{equation}
W_{\alpha\beta} = P^{\beta}_{ij} \partial_j \partial_l G^{\alpha \beta}_{ik}({\bf r}_{\alpha \beta}) P^{\alpha}_{kl},   \tag{B3}
\end{equation}
where ${\bf r}_{\alpha \beta} = {\bf r}_{\beta}-{\bf r}_{\alpha}$ is the separation vector connecting the centers of dipoles $\alpha$ and $\beta$.  Since interaction energy depends on the strain created by one dipole, it also decays as $r^{-2}$, like seen in Fig. 11a.

\typeout{}
\bibliography{bibliography}
\bibliographystyle{ieeetr}

\pagebreak
\clearpage

\onecolumngrid

\section*{Supplementary Information}

\renewcommand{\figurename}{Fig.}
\renewcommand{\thefigure}{S1}
        \begin{figure*}[ht]
            \centering
            \includegraphics[width=12cm]{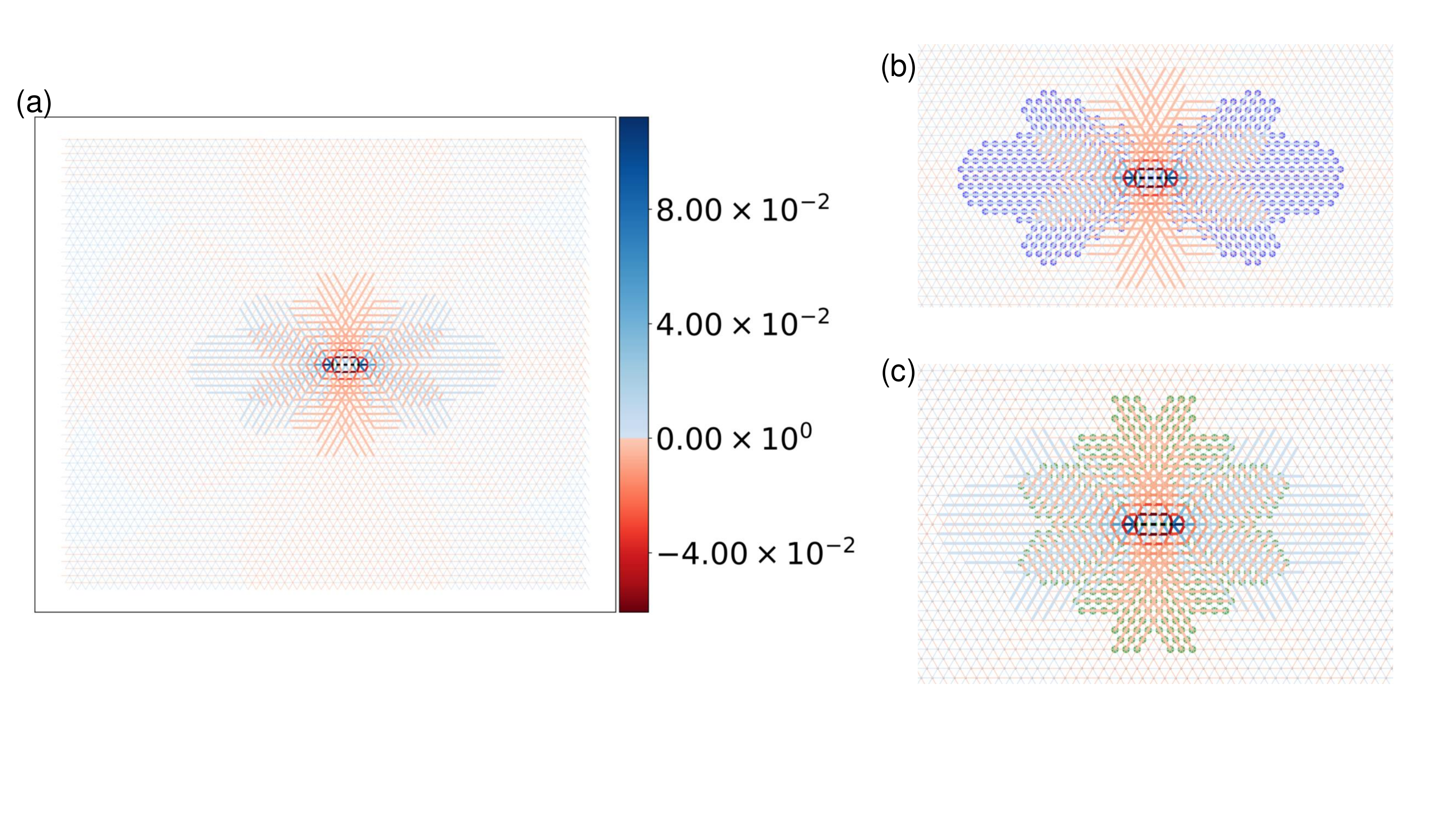}
            \caption{ \textbf{Strain plot and clusters with $\epsilon_{th} = 0.0015$}
            The cluster size increases when we decrease the strain threshold, $\epsilon_{th}$. However, there is no qualitative difference in using other values.}
            \label{fig:cluster small epsilon}
        \end{figure*}


\renewcommand{\thefigure}{S2}
        \begin{figure*}[h]
            \centering
            \includegraphics[width=12cm]{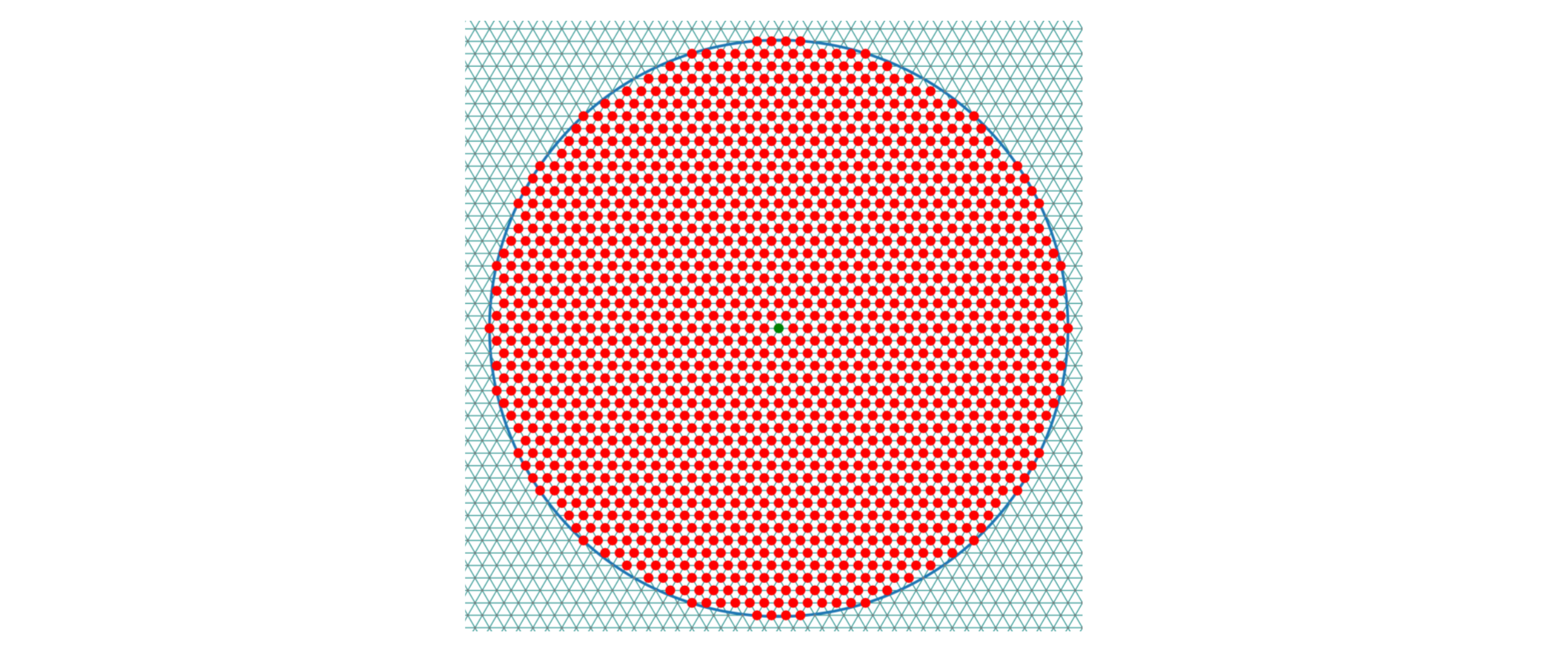}
            \caption{ \textbf{Circular region in triangular network to calculate $R_{g}^{2}/N$.}
            Red nodes are nodes that lie in a circular region of $r = 20$ from the node in green at center. This region is an example of circular regions created to calculate cluster shape parameter $R_{g}^{2}/N$, which for a circle gives a value of 0.1378. Devaition from this value shows asphericity in cluster shapes.}
            \label{fig:Model_circle}
        \end{figure*}


\renewcommand{\thefigure}{S3}
        \begin{figure*}[h]
            \centering
            \includegraphics[width=12cm]{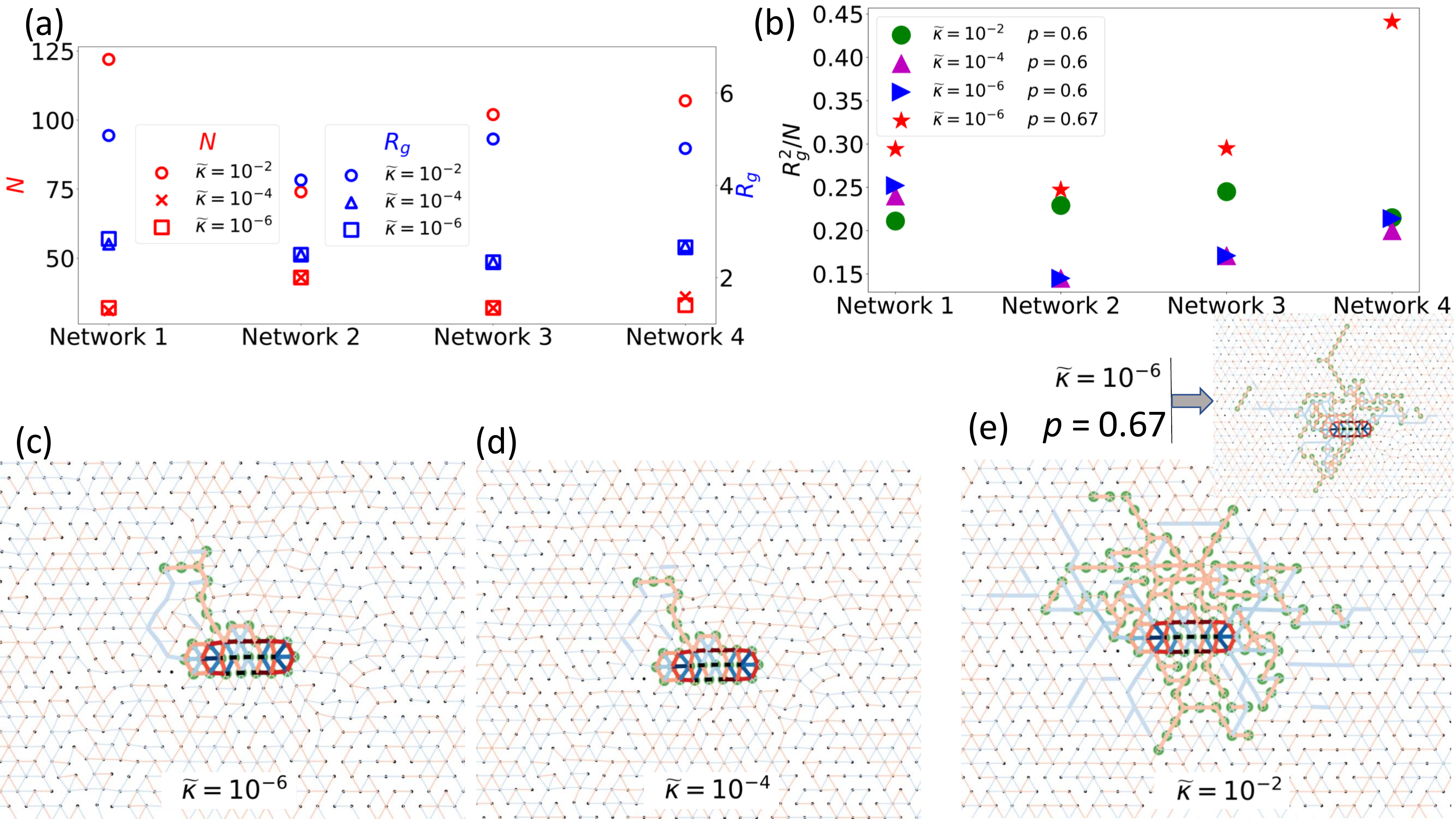}
            \caption{ \textbf{Compressive strain cluster around single dipole in disordered, bending-dominated networks.}
            All results (except inset of e) are for $p = 0.6$, which renders the network to be in the bending regime. (a) Similar to tensile clusters, number of nodes (and Radius of Gyration) in each network shows two distinct regimes: one for $\widetilde{\kappa} = 10^{-6}$ and $10^{-4}$, and another for $\widetilde{\kappa} = 10^{-2}$. (b) For networks 2 and 3, a similar result as (a) is found. However, networks 1 and 4 show that clusters for $p=0.6$ and $\widetilde{\kappa} = 10^{-2}$ have similar values as that for networks with $\widetilde{\kappa} = 10^{-6}$ and $10^{-4}$. This can be attributed to the fact that when clusters are small (for example, c,d show Network 1), a small amount of anisotropicity significantly increases the value of $R_{g}^{2}/N$. Although it is visually evident that the cluster formed when $\widetilde{\kappa} = 10^{-2}$ (d) is much larger and different than those in (c,d), the values of the shape parameter are similar. (e) Inset: The compressive cluster with $p = 0.67$ and $\widetilde{\kappa} = 10^{-6}$ is similar to that of $p = 0.6$ and $\widetilde{\kappa} = 10^{-2}$ - suggesting that these networks lie in the bend-stretch coupled regime. Here the forces percolate farther than for networks in bending regime. This again confirms our previous assertion about accessing the special bend-stretched coupled regime in two ways. }
            \label{fig:compressive cluster bending regime}
        \end{figure*}


\renewcommand{\thefigure}{S4}
        \begin{figure*}[h]
            \centering
            \includegraphics[width=12cm]{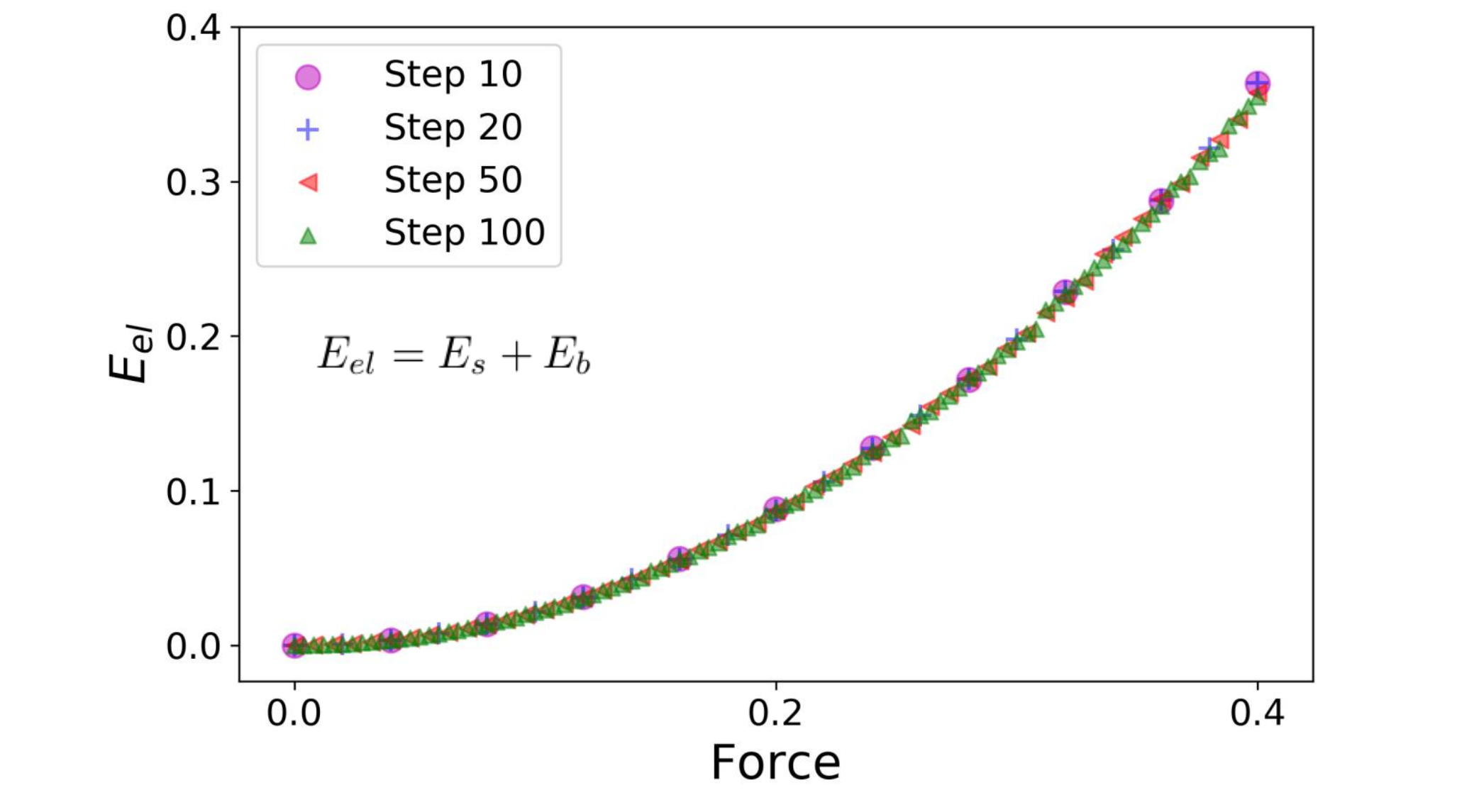}
            \caption{ \textbf{Elastic energy using different force step sizes.}
            Using different force steps, this plot shows the elastic energy of Network 1 with $p = 0.6$ and two dipoles separated along x axis with $d_{x} = 16$. Elastic energy is the sum of stretching and bending energies of the network. The agreement in energy values shows that the simulation is numerically accurate. For all the results in the paper, we took 10 incremental steps of equal $\Delta f = 0.04$ to reach the final force value of $f = 0.4$.}
            \label{fig:energy check}
        \end{figure*}

\renewcommand{\thefigure}{S5}
        \begin{figure*}[h]
            \centering
            \includegraphics[width=12cm]{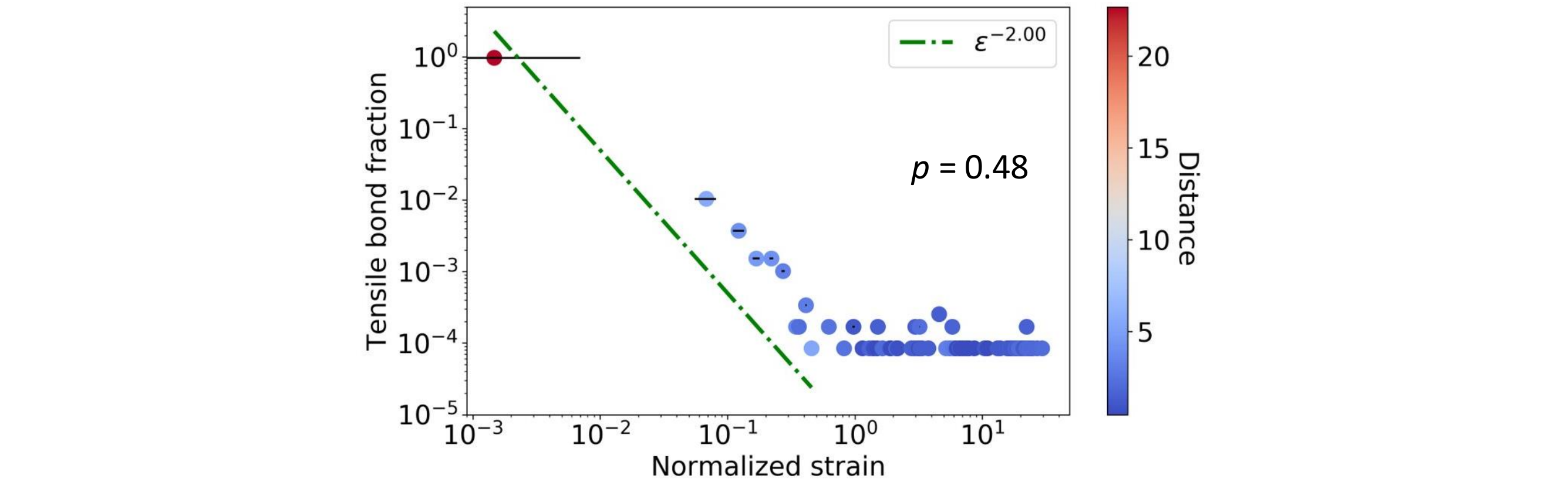}
            \caption{ \textbf{Tensile strain histogram for networks with $p = 0.48$. }
            This plot combines all bonds of the four networks at $p=0.48$. The strain histogram has a gap at intermediate strain values for networks that lie in bending regime.  Almost all tensile bonds are not particularly strained. A very small number of tensile bonds have high strain values.}
            \label{fig:strain energy bend regime}
        \end{figure*}

\renewcommand{\thefigure}{S6}
        \begin{figure*}[h]
            \centering
            \includegraphics[width=12cm]{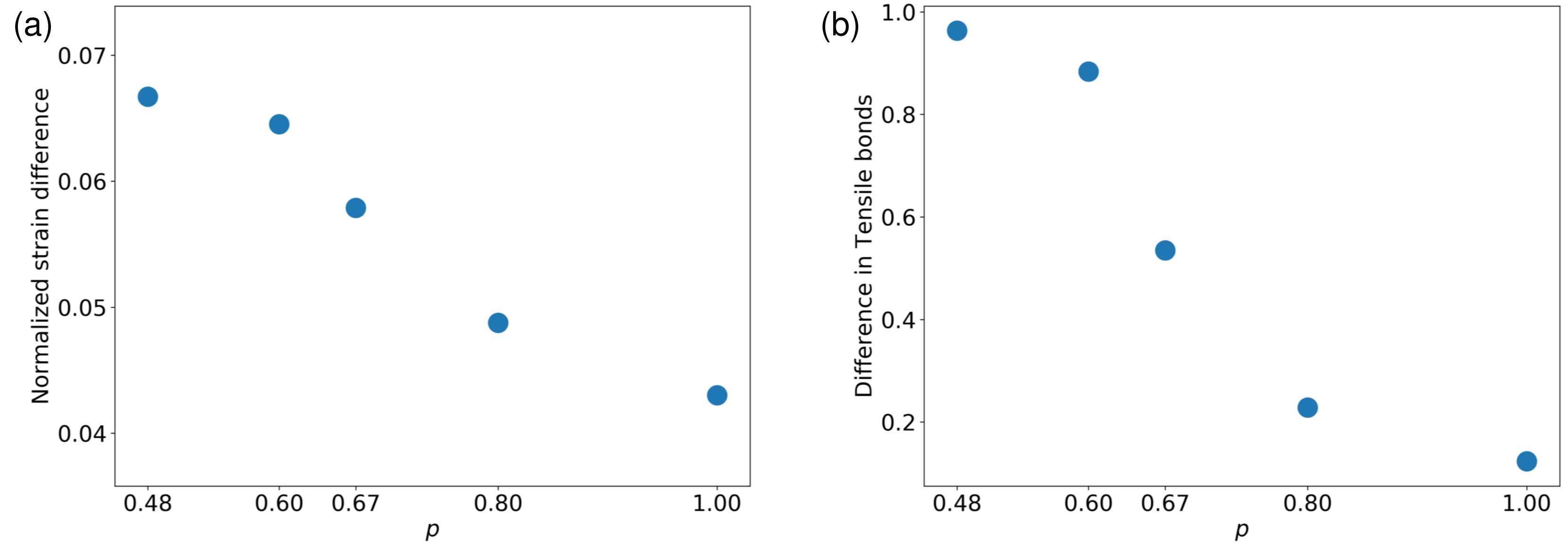}
            \caption{ \textbf{Difference in normalized strains and frequency for first two bins of strain histograms. }
            (a) Difference in normalized strain mean values between the first two bins of histograms in Fig. 7 and SI Fig. 5. The difference increases with dilution. (b) Difference in normalized frequency of the first two bins of histograms in Fig. 7 and SI Fig. 5. Here again, the difference increases with dilution.}
            \label{fig:diff strain energy}
        \end{figure*}

\renewcommand{\thefigure}{S7}
        \begin{figure*}[h]
            \centering
            \includegraphics[width=12cm]{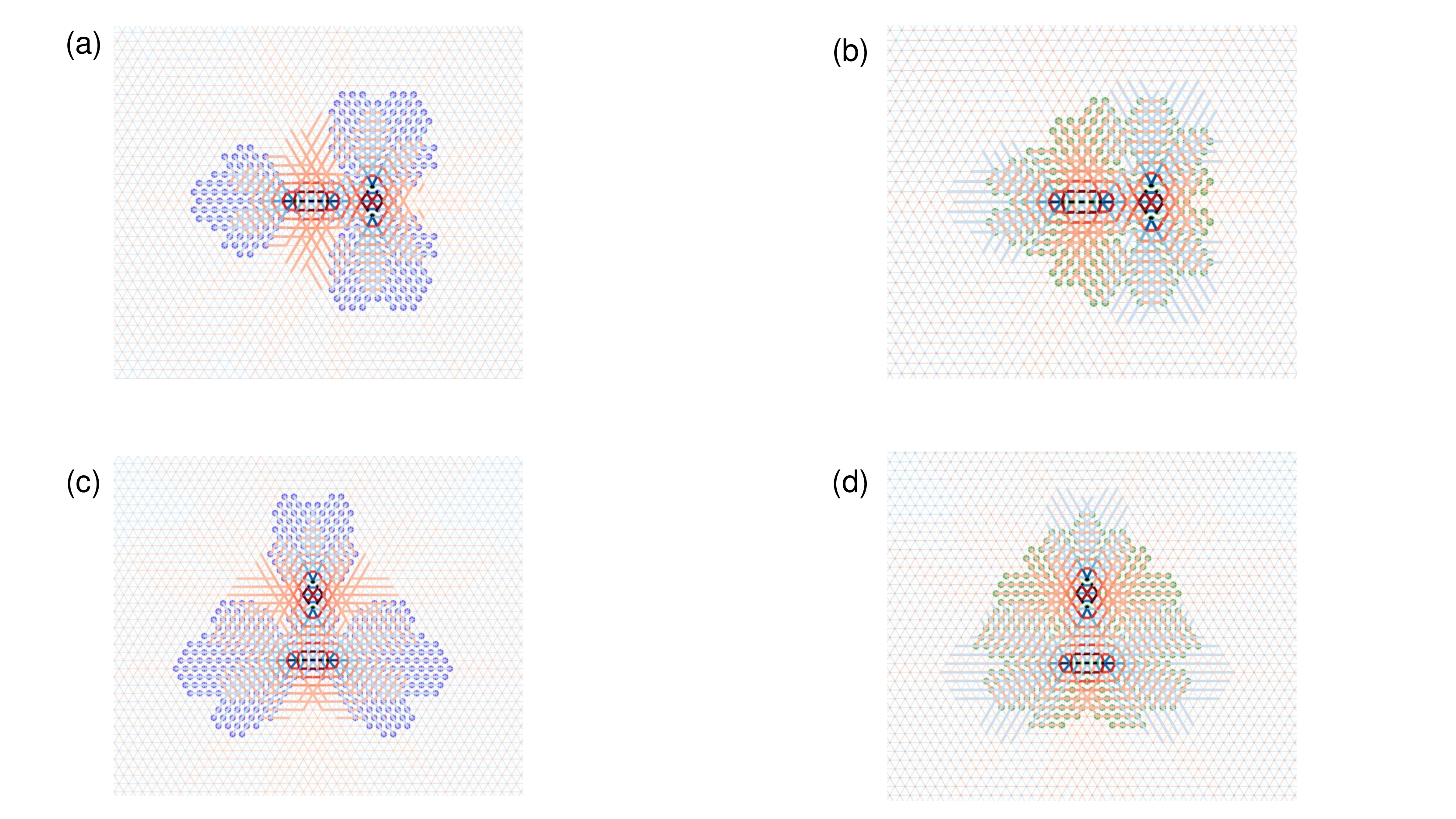}
            \caption{ \textbf{Strain clusters for dipoles that are perpendicular to each other.}
            (a,b) Tensile and compressive clusters for dipoles which are perpendicular to each other and at a distance of $d_{x} = 6$ from each other. (c,d) Tensile and compressive clusters for dipoles that are perpendicular to each at distance of $d_{y} = 8$ rows.}
            \label{fig:perp dipole cluster}
        \end{figure*}

\renewcommand{\thefigure}{S8}
        \begin{figure*}[h]
            \centering
            \includegraphics[width=12cm]{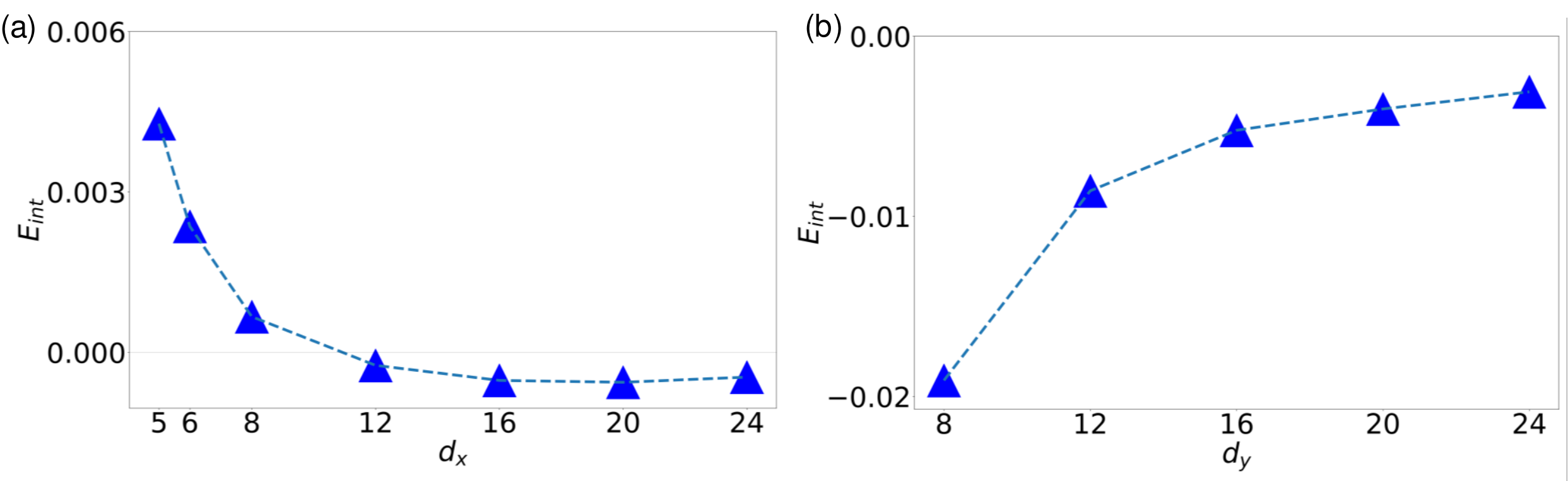}
            \caption{ \textbf{Interaction energy of two dipoles oriented along y axis.}
            (a) The interaction energy of two dipoles oriented along y axis and separated along x axis decreases in magnitude with increasing separation. (b) Two dipoles oriented along y-axis and separated along y axis have a negative interaction energy that decreases in magnitude with increasing separation.}
            \label{fig:int en y orientation}
        \end{figure*}

\end{document}